\documentclass[ALICE,manyauthors]{cernphprep}
\usepackage[comma,square,numbers,sort&compress]{natbib}
\usepackage{hyperref}
\usepackage{lineno}
\usepackage{xspace}
\usepackage{multirow}
\usepackage{orcidlink}
\begin{document}

\newcommand{\pp}              {pp\xspace}
\newcommand{\ppbar}           {\mbox{$\mathrm {p\overline{p}}$}\xspace}
\newcommand{\XeXe}            {\mbox{Xe--Xe}\xspace}
\newcommand{\PbPb}            {\mbox{Pb--Pb}\xspace}
\newcommand{\pA}              {\mbox{pA}\xspace}
\newcommand{\pPb}             {\mbox{p--Pb}\xspace}
\newcommand{\Pbp}             {\mbox{Pb--p}\xspace}
\newcommand{\AuAu}            {\mbox{Au--Au}\xspace}
\newcommand{\dAu}             {\mbox{d--Au}\xspace}

\newcommand{\chiSq}           {\ensuremath{\chi^2}\xspace}
\newcommand{\chiSqNdf}        {\ensuremath{\chi^2 / \mathrm{ndf}}\xspace}
\newcommand{\s}               {\ensuremath{\sqrt{s}}\xspace}
\newcommand{\snn}             {\ensuremath{\sqrt{s_{\mathrm{NN}}}}\xspace}
\newcommand{\pt}              {\ensuremath{p_{\rm T}}\xspace}
\newcommand{\meanpt}          {$\langle p_{\mathrm{T}}\rangle$\xspace}
\newcommand{\ycms}            {\ensuremath{y_{\rm CMS}}\xspace}
\newcommand{\ylab}            {\ensuremath{y_{\rm lab}}\xspace}
\newcommand{\etarange}[1]     {\mbox{$\left | \eta \right | < #1$}}
\newcommand{\yrange}[1]       {\mbox{$\left | y \right | < #1$}}
\newcommand{\etarangeSup}[1]  {\mbox{$\left | \eta \right | > #1$}}
\newcommand{\yrangeSup}[1]    {\mbox{$\left | y \right | > #1$}}
\newcommand{\phirange}[1]     {\mbox{$\left | \phi \right | < #1$}}
\newcommand{\dd}              {\ensuremath{\mathrm{d}}}
\newcommand{\dndy}            {\ensuremath{\mathrm{d}N_\mathrm{ch}/\mathrm{d}y}\xspace}
\newcommand{\dndeta}          {\ensuremath{\mathrm{d}N_\mathrm{ch}/\mathrm{d}\eta}\xspace}
\newcommand{\avdndeta}        {\ensuremath{\langle\dndeta\rangle}\xspace}
\newcommand{\dNdy}            {\ensuremath{\mathrm{d}N_\mathrm{ch}/\mathrm{d}y}\xspace}
\newcommand{\Npart}           {\ensuremath{N_\mathrm{part}}\xspace}
\newcommand{\avNpart}         {\ensuremath{\langle\Npart\rangle}\xspace}
\newcommand{\NpartMult}       {\ensuremath{N^\mathrm{mult}_\mathrm{part}}\xspace}
\newcommand{\avNpartMult}     {\ensuremath{\langle\NpartMult\rangle}\xspace}
\newcommand{\Ncoll}           {\ensuremath{N_\mathrm{coll}}\xspace}
\newcommand{\avNcoll}         {\ensuremath{\langle\Ncoll\rangle}\xspace}
\newcommand{\NcollMult}       {\ensuremath{N^\mathrm{mult}_\mathrm{coll}}\xspace}
\newcommand{\avNcollMult}     {\ensuremath{\langle\NcollMult\rangle}\xspace}
\newcommand{\Taa}             {\ensuremath{T_{\rm AA}}\xspace}
\newcommand{\avTaa}           {\ensuremath{\langle T_{\rm AA} \rangle}\xspace}
\newcommand{\dEdx}            {\ensuremath{\textrm{d}E/\textrm{d}x}\xspace}
\newcommand{\RpPb}            {\ensuremath{R_{\rm pPb}}\xspace}
\newcommand{\RPbPb}           {\ensuremath{R_{\rm AA}}\xspace}

\newcommand{\nineH}           {$\sqrt{s}~=~0.9$~Te\kern-.1emV\xspace}
\newcommand{\seven}           {$\sqrt{s}~=~7$~Te\kern-.1emV\xspace}
\newcommand{\twoH}            {$\sqrt{s}~=~0.2$~Te\kern-.1emV\xspace}
\newcommand{\twosevensix}     {$\sqrt{s}~=~2.76$~Te\kern-.1emV\xspace}
\newcommand{\five}            {$\sqrt{s}~=~5.02$~Te\kern-.1emV\xspace}
\newcommand{\twosevensixnn}   {$\sqrt{s_{\mathrm{NN}}}~=~2.76$~Te\kern-.1emV\xspace}
\newcommand{\fivenn}          {$\sqrt{s_{\mathrm{NN}}}~=~5.02$~Te\kern-.1emV\xspace}
\newcommand{\eightnn}         {$\sqrt{s_{\mathrm{NN}}}~=~8.16$~Te\kern-.1emV\xspace}
\newcommand{\LT}              {L{\'e}vy-Tsallis\xspace}
\newcommand{\TeVc}            {Te\kern-.1emV$/c$\xspace}
\newcommand{\GeVc}            {Ge\kern-.1emV$/c$\xspace}
\newcommand{\MeVc}            {Me\kern-.1emV$/c$\xspace}
\newcommand{\TeV}             {Te\kern-.1emV\xspace}
\newcommand{\GeV}             {Ge\kern-.1emV\xspace}
\newcommand{\MeV}             {Me\kern-.1emV\xspace}
\newcommand{\GeVmass}         {Ge\kern-.2emV$/c^2$\xspace}
\newcommand{\MeVmass}         {Me\kern-.2emV$/c^2$\xspace}
\newcommand{\lumi}            {\ensuremath{\mathcal{L}}\xspace}
\newcommand{\intlumi}         {\ensuremath{\lumi_\mathrm{int}}}

\newcommand{\ITS}             {\rm{ITS}\xspace}
\newcommand{\TOF}             {\rm{TOF}\xspace}
\newcommand{\ZDC}             {\rm{ZDC}\xspace}
\newcommand{\ZDCs}            {\rm{ZDCs}\xspace}
\newcommand{\ZNA}             {\rm{ZNA}\xspace}
\newcommand{\ZNC}             {\rm{ZNC}\xspace}
\newcommand{\SPD}             {\rm{SPD}\xspace}
\newcommand{\SDD}             {\rm{SDD}\xspace}
\newcommand{\SSD}             {\rm{SSD}\xspace}
\newcommand{\TPC}             {\rm{TPC}\xspace}
\newcommand{\TRD}             {\rm{TRD}\xspace}
\newcommand{\VZERO}           {\rm{V0}\xspace}
\newcommand{\VZEROA}          {\rm{V0A}\xspace}
\newcommand{\VZEROC}          {\rm{V0C}\xspace}
\newcommand{\Vdecay} 	      {\ensuremath{V^{0}}\xspace}

\newcommand{\ee}           {\ensuremath{e^{+}e^{-}}}
\newcommand{\pip}          {\ensuremath{\pi^{+}}\xspace}
\newcommand{\pim}          {\ensuremath{\pi^{-}}\xspace}
\newcommand{\kap}          {\ensuremath{\rm{K}^{+}}\xspace}
\newcommand{\kam}          {\ensuremath{\rm{K}^{-}}\xspace}
\newcommand{\pbar}         {\ensuremath{\rm\overline{p}}\xspace}
\newcommand{\kzero}        {\ensuremath{{\rm K}^{0}_{\rm{S}}}\xspace}
\newcommand{\lmb}          {\ensuremath{\Lambda}\xspace}
\newcommand{\almb}         {\ensuremath{\overline{\Lambda}}\xspace}
\newcommand{\Om}           {\ensuremath{\Omega^-}\xspace}
\newcommand{\Mo}           {\ensuremath{\overline{\Omega}^+}\xspace}
\newcommand{\X}            {\ensuremath{\Xi^-}\xspace}
\newcommand{\Ix}           {\ensuremath{\overline{\Xi}^+}\xspace}
\newcommand{\Xis}          {\ensuremath{\Xi^{\pm}}\xspace}
\newcommand{\Oms}          {\ensuremath{\Omega^{\pm}}\xspace}
\newcommand{\W}            {\ensuremath{\rm{W}^\pm}\xspace}
\newcommand{\Wminus}       {\ensuremath{\rm{W}^-}\xspace}
\newcommand{\Wplus}        {\ensuremath{\rm{W}^+}\xspace}
\newcommand{\Z}            {\ensuremath{\rm{Z}^0}\xspace}
\newcommand{\jpsi}         {\ensuremath{\rm{J}/\psi}\xspace}
\newcommand{\upsi}         {\ensuremath{\Upsilon}\xspace}
\newcommand{\muon}         {\ensuremath{\mu^-}\xspace}
\newcommand{\Amuon}        {\ensuremath{\mu^+}\xspace}
\newcommand{\muonpm}       {\ensuremath{\mu^\pm}\xspace}
\newcommand{\num}          {\ensuremath{\nu_\mu}\xspace}
\newcommand{\Anum}         {\ensuremath{\overline{\nu}_\mu}\xspace}

\begin{titlepage}
\PHyear{2022}       
\PHnumber{076}      
\PHdate{30 March}  

\title{W$\mathbf{^\pm}$-boson production in \pPb collisions at $\mathbf{\sqrt{\mathit{s}_{NN}}}$~=~8.16~Te\kern-.1emV\xspace \\ and \PbPb collisions at $\mathbf{\sqrt{\mathit{s}_{NN}}}$~=~5.02~Te\kern-.1emV\xspace}
\ShortTitle{\W bosons in p--Pb at \eightnn and Pb--Pb at \fivenn}   

\Collaboration{ALICE Collaboration\thanks{See Appendix~\ref{app:collab} for the list of collaboration members}}
\ShortAuthor{ALICE Collaboration} 

\begin{abstract}
The production of the \W bosons measured in \pPb collisions at a centre-of-mass energy per nucleon--nucleon collision \eightnn and \PbPb collisions at \fivenn with ALICE at the LHC is presented. The \W bosons are measured via their muonic decay channel, with the muon reconstructed in the pseudorapidity region $-4 < \eta^\mu_{\rm lab} < -2.5$ with transverse momentum $\pt^\mu~>$~10~\GeVc. While in \PbPb collisions the measurements are performed in the forward ($2.5 < y^\mu_{\rm cms} < 4$) rapidity region, in \pPb collisions, where the centre-of-mass frame is boosted with respect to the laboratory frame, the measurements are performed in the backward ($-4.46 < y^\mu_{\rm cms} < -2.96$) and forward ($2.03 < y^\mu_{\rm cms} < 3.53$) rapidity regions. The \Wminus and \Wplus production cross sections, lepton-charge asymmetry, and nuclear modification factors are evaluated as a function of the muon rapidity. In order to study the production as a function of the \pPb collision centrality, the production cross sections of the \Wminus and \Wplus bosons are combined and normalised to the average number of binary nucleon--nucleon collision \avNcoll. In \PbPb collisions, the same measurements are presented as a function of the collision centrality. Study of the binary scaling of the \W-boson cross sections in \pPb and \PbPb collisions is also reported. The results are compared with perturbative QCD calculations, with and without nuclear modifications of the Parton Distribution Functions (PDFs), as well as with available data at the LHC. Significant deviations from the theory expectations are found in the two collision systems, indicating that the measurements can provide additional constraints for the determination of nuclear PDFs and in particular of the light-quark distributions.
\end{abstract}
\end{titlepage}

\setcounter{page}{2} 


\section{Introduction}
\label{sec:intro}

The production of the \W- and \Z-vector bosons is extensively studied at hadron colliders. The \W and \Z bosons are weakly interacting particles, produced early in hadronic collisions (with a formation time $t_{\rm f} \sim 1/M \sim 10^{-3}$~fm$/c$), predominantly via the Drell-Yan process in which a quark--antiquark pair annihilates into a lepton pair~\cite{drell-yan, drell-yan-n3ll}. Due to their large masses, $M_{\W} = 80.379 \pm 0.012$~\GeVmass and $M_{\Z} = 91.1876 \pm 0.0021$~\GeVmass~\cite{pdg}, their production is well described within the perturbative quantum chromodynamics (pQCD) framework, up to Next-to-Next-to-Leading Order (NNLO) by means of the QCD factorisation theorem for hard processes~\cite{vecbos-nnlo-1, vecbos-nnlo-2}. Factorisation allows us to separate the short distance part of the cross section, corresponding to the partonic cross section that can be expanded perturbatively, from the long distance part containing the Parton Distribution Functions (PDFs), parameterising the partonic content of the nucleon and determined from experimental data. The input parameters for theoretical calculations, such as the boson masses or the weak couplings, are known with high accuracy, enabling the usage of measurements of the electroweak-boson production to determine the up (u), down (d) and to a lesser extent strange (s) PDFs (see Refs.~\cite{pdf-review, hadron-struct} for recent reviews). In nuclear collisions, the presence of a nuclear environment affects the inner structure of the nucleon, requiring the determination of nuclear PDFs (nPDFs). As for the free-nucleon case, the nPDFs are obtained from a global analysis of the available data, but in this case the results are mostly constrained by Deep-Inelastic Scatterings (DIS) and Drell-Yan data in a limited region of the four-momentum transfer squared $Q^2$ and parton longitudinal momentum fraction $x$ (Bjorken-$x$). The resulting nPDF uncertainties drastically limit the precision of theoretical calculations and their ability to describe and predict processes in nuclear collisions. In order to further constrain the nPDFs and reduce their uncertainties, the production of the \W and \Z bosons has been measured in proton--lead (\pPb) and lead--lead (\PbPb) collisions at the CERN Large Hadron Collider (LHC) by the four main experiments, at midrapidity by ATLAS and CMS~\cite{Atlas:ZpPb5tev, Atlas:WPbPb2tev, Atlas:ZPbPb2tev, Atlas:WPbPb5tev, Atlas:ZPbPb5tev, Cms:WpPb5tev, Cms:ZpPb5tev, Cms:WpPb8tev, Cms:WPbPb2tev, Cms:ZPbPb2tev, Cms:ZPbPb2tev-2, Cms:ZPbPb5tev} and at large rapidities by ALICE and LHCb~\cite{Alice:WZpPb5tev, Alice:ZPbPb2tev, Alice:ZpPb8tevPbPb5tev, Lhcb:ZpPb5tev}.

Four main intervals of Bjorken-$x$ featuring different nuclear modifications can be distinguished at high $Q^2$ values. The nPDFs show a suppression at low Bjorken-$x$, for $x \lesssim 0.05$, and an enhancement within the range $x \sim 0.05 - 0.3$\footnote{All the Bjorken-$x$ ranges are indicative, as the precise values of the region boundaries depend on the parton flavour, the nPDF parametrisation, and the $Q^2$ scale.}. Both these effects, referred to as shadowing and anti-shadowing, respectively, originate from destructive or constructive interferences of amplitudes arising from multiple scatterings between partons in the nucleus~\cite{shadowing}. Another depletion region is seen for $x$ within 0.3 -- 0.9 in the so-called EMC-effect region which is not yet fully understood~\cite{emc}. Finally, for $x$ larger than 0.9 the Fermi motion of the nucleons inside the nucleus yields an enhancement of the PDF~\cite{fermi}. These effects will naturally affect the production of electroweak bosons~\cite{shadowing-wkbos}, and their measurement provides a unique opportunity to constrain the nPDFs at high $Q^2 \sim M^2_{\rm W,Z}$. Moreover, with the large luminosities and centre-of-mass energies delivered by the LHC, combined with the wide acceptance covered by the LHC experiments, the study of electroweak bosons has become accessible in \pPb and \PbPb collisions over a large Bjorken-$x$ range, from almost unity down to $x \sim 10^{-4}$ where the experimental constraints are scarce. Measurements in \pPb collisions at large negative and positive rapidities are of high interest as they allow the disentanglement of the high ($\sim 10^{-1}$) and low ($\sim 10^{-4} - 10^{-3}$) Bjorken-$x$ intervals, respectively. The yields of the \Wminus and \Wplus bosons, mainly produced by interactions between u and d quarks via the $\mathrm{d} \overline{\mathrm{u}} \rightarrow \Wminus$ and $\mathrm{u} \overline{\mathrm{d}} \rightarrow \Wplus$ processes, offer a probe of the light quark PDFs, while their asymmetry is sensitive to the down-to-up ratio in the nucleus~\cite{ZWnPDFconstraints}. The leptonic decay of these bosons is of particular interest, as the decay products do not interact strongly, therefore being blind to the quark--gluon plasma (QGP), the hot and dense medium created in heavy-ion collisions. In addition, the in-medium energy loss of the decay leptons by bremsstrahlung is negligible~\cite{bremsstrahlung}. Combined with the colourless nature of the \W boson itself, this physics channel provides a medium-blind process and consequently, a direct probe of the initial state of the collision even in the presence of a QGP. The production of electroweak bosons, therefore, enables the study of the nPDFs of the colliding nuclei.

The measurements of the \W-boson production presented in this publication are compared with predictions obtained from calculations at Next-to-Leading Order (NLO), implementing the nuclear modifications of the PDFs using the EPPS16~\cite{epps16}, nCTEQ15~\cite{ncteq15} and nNNPDF2.0~\cite{nnnpdf} sets, in which the parametrisation and determination of the nPDF follow different approaches. The approach of the EPPS (formerly EPS) group introduces, for a given parton $i$ in a nucleus with atomic number $A$, the nuclear correction factor $R_i (x,A)$ at the input parametrisation scale $Q^2_0$. In such a model, the nPDF set is composed of nuclear modification functions to be applied to a free-nucleon PDF set which serves as a baseline. The approach of the nCTEQ collaboration does not utilise the nuclear correction factors, instead, it is a full nPDF parametrisation. It starts from the functional form used for the free-proton PDF (in the nCTEQ case the form is similar to the CTEQ6 parametrisation~\cite{cteq6}), with the addition of $A$-dependent free parameters. The lack of experimental data that can be used for the nPDF determination induces a strong dependence of the models on the phenomenological and methodological assumptions. The EPPS16 and nCTEQ15 sets show large differences in the predicted nuclear modifications and associated uncertainties~\cite{npdftoday}, originating from the functional form, the number of free parameters, and the data points included in the global analysis. In order to reduce the parametrisation bias, the nNNPDF collaboration adopted the methodology described in Ref.~\cite{nnpdf}, and used artificial neural networks as universal, unbiased interpolants to parametrise the $x$ and $A$ dependence of the nPDFs. Recently, the LHC experiments contributed to the evolution of the models, and \W and \Z measurements in \pPb collisions are now included into the input datasets, in EPPS starting with EPPS16~\cite{epps16}, in nCTEQ after the nCTEQ15WZ update~\cite{ncteq15wz}, and in nNNPDF from their 2.0 release~\cite{nnnpdf}. It should be noted that the EPPS model has recently been updated with the release of the EPPS21 set~\cite{epps21}. The production of electroweak bosons calculated from this set is in fair agreement with the ones obtained with the EPPS16 model, with a significant reduction of the associated uncertainties.

In this article, the ALICE results on the measurement of the \W-boson production via the muonic decay channel in \pPb collisions at a centre-of-mass energy per nucleon--nucleon collision \eightnn and \PbPb collisions at \fivenn are reported. These results constitute the first measurements of the \W-boson production at large rapidities for these collision systems and energies, with the p--Pb results complementing the CMS measurements at midrapidity~\cite{Cms:WpPb8tev} and extending the ALICE measurements in \pPb collisions at \fivenn~\cite{Alice:WZpPb5tev}. The paper is structured as discussed in the following. Section~\ref{sec:data} introduces the ALICE apparatus, focusing on the detectors relevant for the analyses, followed by a description of the event and track selections. The analysis strategy, including the procedure for the signal extraction and the simulation of the apparatus, is presented in Section~\ref{sec:analysis}, together with a discussion of the systematic uncertainties. The results are reported in Section~\ref{sec:results} where they are compared with theoretical predictions and other published measurements. A summary of the results and their interpretation is given in Section~\ref{sec:summary}.

\section{ALICE apparatus and data samples}
\label{sec:data}

  \subsection{The ALICE detector}

The \W bosons are detected through their muonic decay channel via the $\Wminus \rightarrow \muon \Anum$ process and its charge conjugate, with a branching ratio BR$~=~(10.63 \pm 0.15)$\%~\cite{pdg}, from data recorded with the ALICE muon spectrometer~\cite{tdr-muon, tdr-muon-addendum}. The spectrometer covers in full azimuth the $-4 < \eta_{\rm lab} < -2.5$ pseudorapidity interval\footnote{In the ALICE reference frame, the muon spectrometer covers negative $\eta$. In symmetric collisions such as \PbPb, positive values of rapidity are conventionally used for the muon coverage. In \pPb collisions, by convention, the proton beam moves towards positive rapidities.}. Its tracking system is composed of five stations, each made of two planes of cathode pad chambers. The third station sits inside a dipole magnet providing an invertible magnetic field with integrated intensity of 3 Tm, which bends the trajectory of charged particles thus enabling the measurement of the track momentum. The muon system also includes a muon trigger, consisting of four planes of resistive plate chambers arranged in two stations. The whole spectrometer is shielded by a set of absorbers. A conical absorber of 10 interaction lengths ($\lambda_i$) made of carbon, concrete, and steel is located in front of the muon spectrometer, filtering out hadrons and low-momentum muons from the decays of light particles such as pions and kaons. The trigger stations are located behind a 1.2 m thick (about 7.2 $\lambda_i$) iron wall, absorbing hadrons punching through the front absorber as well as low-momentum secondary muons. Finally, a high-density cylinder made of tungsten and lead, the so-called small-angle absorber, surrounds the beam pipe throughout the muon spectrometer in its entirety and shields it against secondary particles produced by the interaction of primary particles at large $\eta$ with the beam pipe.

Other detectors are needed for primary vertex reconstruction, triggering on Minimum Bias (MB) collisions, multiplicity determination, and centrality evaluation. The primary interaction vertex reconstruction is performed using the Silicon Pixel Detector (SPD), the two innermost layers of the Inner Tracking System (ITS)~\cite{tdr-its}, covering the pseudorapidity intervals $|\eta_{\rm lab}| < 2.0$ and $|\eta_{\rm lab}| < 1.4$. The V0 detector~\cite{tdr-forward} is made of two arrays of scintillator tiles, located asymmetrically around the collision point, along the beam direction, at $z = 3.4$ m (V0A) and $z = -0.9$ m (V0C), and covering the pseudorapidity intervals $2.8 < \eta_{\rm lab} < 5.1$ and $-3.7 < \eta_{\rm lab} <-1.7$, respectively. The V0 provides an online MB trigger through the logical coincidence of a signal in the two arrays, and participates in the determination of the luminosity by providing a reference process for van der Meer scans~\cite{vanderMeer}. It is also used for the evaluation of the centrality in Pb--Pb collisions by means of a Glauber model fit~\cite{glauber, denterria2021} to the sum of the signal amplitudes in the two arrays (the V0M estimator). This allows one to classify the events in centrality classes corresponding to a percentile of the total hadronic cross section. The centrality evaluated in this way relies on the event charged-particle multiplicity, a method which has been shown to be strongly biased in p--Pb collisions~\cite{centpPb5tev}. Instead, the centrality estimation for this system uses the Zero Degree Calorimeter (ZDC)~\cite{tdr-zdc}, a set of two hadronic calorimeters located along the beam pipe, on both sides of the collision point, 112.5 m away from it. The timing information delivered by the V0 and ZDC detectors also helps to reduce the beam-induced background. A complete description of the ALICE detector can be found in Ref.~\cite{alice} and its performance is reported in Ref.~\cite{alice_perf3}, where standard detection, reconstruction, and analysis procedures are described.

  \subsection{Event and track selections}
    \label{sec:selection}

The analysis in \pPb collisions uses the data samples collected in 2016 at \eightnn. These data were taken in two colliding beam configurations, with either the protons or lead ions moving towards the spectrometer, hereafter referred to as the p-going and Pb-going configurations, respectively. By convention, the protons move towards positive rapidities. Because of the single magnet design of the LHC, the proton and Pb beams have the same magnetic rigidity, leading to different energies per nucleon, amounting to 6.5 TeV for the protons and 2.56 TeV for the Pb ions. The resulting nucleon--nucleon centre-of-mass system is thus boosted with respect to the laboratory frame, resulting in a rapidity shift of $\Delta y_{\rm cms/lab} = 0.465$ in the direction of the proton beam. The rapidity acceptance of the spectrometer in the centre-of-mass system is then $2.03 < y_{\rm cms} < 3.53$ in the p-going direction and $-4.46 < y_{\rm cms} < -2.96$ in the Pb-going one. The analysis in \PbPb collisions uses the data samples collected in 2015 and 2018 at \fivenn in the rapidity range $2.5 < y^\mu_{\rm cms} < 4$. For each sample, two sub-periods can be distinguished according to the sign of the magnetic field delivered by the dipole magnet.

The analysed data samples consist of events with at least one muon track candidate selected by the muon trigger system, with an online selection on the transverse momentum ($\pt^\mu$) requiring it to be above $\simeq$~4.2~\GeVc (at the threshold, the track produces a trigger signal with a 50\% probability), in coincidence with a MB signal in the V0 detector. The \PbPb analysis is limited to the most central 90\% of the total hadronic cross section, where the MB trigger is fully efficient and electromagnetic interactions are negligible. The events were further required to have a reconstructed vertex position along the beam direction within $\pm 10$ cm from the nominal interaction point in order to keep the full efficiency of the SPD for vertex reconstruction. Events in which two or more interactions occur in the same colliding bunch (in-bunch pile-up) or during the readout time of the SPD (out-of-bunch pile-up), amounting to about 20\% of the sample, are removed using the information from the SPD and V0 detectors. The integrated luminosity was evaluated by estimating the equivalent number of MB events corresponding to the muon-triggered data sample and then dividing by $\sigma_{\rm V0M}$, the V0 visible cross section measured by means of van der Meer scans~\cite{vanderMeer,lumi-pPb8tev,lumi-PbPb5tev}. The number of MB events corresponding to the muon-triggered sample was evaluated as $N_{\rm MB} = F_{\mu\text{-trig} / \text{MB}} \times N_{\mu\text{-trig}}$, where $N_{\mu\text{-trig}}$ is the number of muon-triggered events and $F_{\mu\text{-trig} / \text{MB}}$ is the inverse of the probability to have a muon trigger in a MB event. The value of the normalisation factor $F_{\mu\text{-trig} / \text{MB}}$ was evaluated with two different methods, either by applying the muon trigger condition in the analysis of MB events, or by comparing the counting rate of the two triggers, both corrected for pile-up effects. The nominal value was obtained from the method using the trigger rates, while the difference between the two methods was taken as the systematic uncertainty on the normalisation factor. This uncertainty amounts to 1.4\% (1.1\%) in \pPb collisions for the p-going (Pb-going) configuration, and to 1\% in \PbPb collisions. The integrated luminosities of the considered \pPb data samples amount to $6.73 \pm 0.16$~nb$^{-1}$ and $10.0 \pm 0.22$~nb$^{-1}$ in the p-going and Pb-going directions, respectively, and to $663 \pm 15$ $\mu$b$^{-1}$ for \PbPb collisions after merging the 2015 and 2018 data samples. The quoted uncertainties are the systematic uncertainties, while the statistical ones are negligible.

The classification of the events in \pPb collisions into centrality intervals is performed based on the energy deposited in the neutron calorimeters (ZN) of the ZDC in the direction of the Pb fragments. For each of these intervals, the average number of binary nucleon--nucleon collisions \avNcoll is obtained from the hybrid method described in Ref.~\cite{centpPb5tev}. The method relies on the assumption that the charged-particle multiplicity measured at midrapidity is proportional to the average number of nucleons participating in the interaction \avNpart. The values of \avNpart for a given ZN-centrality class are calculated by scaling the average number of participants in MB collisions $\langle N^{\rm MB}_{\rm part} \rangle$, estimated by means of Glauber Monte Carlo (MC)~\cite{glauber-improved,centrality}, with the ratio of the average charged-particle multiplicity measured at midrapidity for the ZN-centrality class to that in MB collisions. In the following, these values are denoted \avNpartMult to indicate this assumption. The corresponding number of binary collisions is then obtained as $\avNcollMult = \langle N^{\rm mult}_{\rm part} \rangle - 1$. The associated uncertainty is evaluated using different approaches as described in Ref.~\cite{centrality}. The resulting values of \avNcollMult and their uncertainties are summarised in Table~\ref{table:centrality-pPb}. In \PbPb collisions, the centrality is determined from the distribution of the signal amplitude in the V0 arrays and is expressed in percentages of the total hadronic cross section. The collisional geometrical properties \avNpart, \avNcoll, and the nuclear overlap function \avTaa of the different centrality intervals are obtained via a Glauber model fit to the V0 signal amplitude distribution. The Glauber model is also used to determine the so-called anchor point below which the centrality determination is not reliable. The values of \avTaa in \PbPb collisions at \fivenn are given in Table~\ref{table:centrality-PbPb} for the centrality classes considered in this work.

\begin{table}
    \centering
    \caption{Average number of binary nucleon--nucleon collisions \avNcollMult estimated with the hybrid method for the ZN centrality classes in \pPb collisions at \eightnn~\cite{centrality}.}
    \begin{tabular}{|c||c||c|c|c|c|}
      \hline
      Centrality class & 0--100\%        & 0--20\%         & 20--40\%        & 40--60\%        & 60--100\% \\
      \hline
      \avNcollMult     & $7.09 \pm 0.28$ & $12.2 \pm 0.52$ & $9.81 \pm 0.17$ & $7.09 \pm 0.29$ & $3.17 \pm 0.09$ \\
      \hline
    \end{tabular}
    \label{table:centrality-pPb}
\end{table}

\begin{table}
    \centering
    \caption{Average nuclear overlap function \avTaa evaluated with a Glauber MC fit to the sum of the V0 amplitudes in \PbPb collisions at \fivenn~\cite{centrality}.}
    \begin{tabular}{|c||c||c|c|c|c|}
      \hline
      Centrality class                & 0--90\%         & 0--10\%          & 10--20\%         & 20--40\%        & 40--90\% \\
      \hline
      \avTaa (mb\textsuperscript{-1}) & $6.28 \pm 0.06$ & $23.26 \pm 0.17$ & $14.40 \pm 0.13$ & $6.93 \pm 0.09$ & $1.00 \pm 0.02$ \\
      \hline
    \end{tabular}
    \label{table:centrality-PbPb}
\end{table}

The muon track candidates reconstructed in the events passing the requirements described above are selected according to the following criteria. A fiducial selection is applied on the track pseudorapidity, requiring it to be in the interval $-4 < \eta^\mu_{\rm lab} < -2.5$ to remove the particles at the edge of the spectrometer acceptance. An additional selection on the polar angle measured at the end of the front absorber, of $170^\circ < \theta_{\rm abs} < 178^\circ$, rejects the tracks crossing the high-density region of the front absorber, where they experience significant multiple scatterings. The contamination by tracks not pointing to the nominal interaction vertex, mostly originating from beam--gas interactions and secondary particles produced in the front absorber, is efficiently removed by exploiting the correlation between the track momentum $p$ and its distance of closest approach (DCA) to the vertex (i.e., the distance to the primary vertex of the track trajectory projected on the plane transverse to the beam axis). Being subject to multiple scatterings in the front absorber, the DCA of particles produced in the collision follows a Gaussian distribution, with a sigma depending on the material crossed and being proportional to the inverse of the momentum $p$. Background tracks, on the other hand, have on average a DCA larger than about 40 cm, independently of their momentum. A selection on the product of the track momentum with its DCA ($p \times$DCA) allows the suppression of this background source down to a negligible level. Finally, the muon identification is performed by matching the track reconstructed in the tracking system with a track segment in the trigger stations. The track in the tracking system is extrapolated to the trigger stations, and a $\chi^2$-based criterion determines the quality of the matching.

\section{Analysis strategy}
\label{sec:analysis}

  \subsection{Overview}

The \W bosons are detected through their muonic decay channel via the $\Wminus \rightarrow \muon\Anum$ and $\Wplus \rightarrow \Amuon\num$ processes following the method described in Ref.~\cite{Alice:WZpPb5tev}. Since ALICE is not a hermetic detector, one cannot reconstruct the missing transverse energy due to the presence of a neutrino in the final state. The signal extraction is therefore performed from the single muon \pt distribution, excluding the $\pt^\mu~<$~10~\GeVc interval where the signal-to-background ratio is very small. One can distinguish three main contributions to the inclusive spectrum, namely muons originating from the decay of \W, \Z/$\gamma^*$, and heavy-flavour (charm and beauty) hadrons. The signal extraction procedure relies on templates, which are generated by means of MC simulations, and are used to fit the measured muon \pt distributions according to
\begin{equation}
  f(\pt) = N^{\rm raw}_{\rm HF} f_{\rm HF} (\pt) + N^{\rm raw}_{\muonpm \leftarrow \W} \left( f_{\muonpm \leftarrow \W} (\pt) + R \times f_{\muonpm \leftarrow \Z / \gamma^*} (\pt) \right),
  \label{eqn:signal_extraction}
\end{equation}
where $f_{\rm HF}$, $f_{\muonpm \leftarrow \W}$, and $f_{\muonpm \leftarrow \Z/\gamma^*}$ are the templates accounting for muons from heavy-flavour hadrons, \W-boson, and \Z/$\gamma^*$ decays, respectively. The number of muons from heavy-flavour hadrons and \W-boson decays ($N^{\rm raw}_{\rm HF}$ and $N^{\rm raw}_{\muonpm \leftarrow \W}$) are free parameters of the fit, while the number of muons from \Z/$\gamma^*$ decays is forced to be proportional to that of \W decays according to the ratio $R$ of their production cross sections as predicted by MC simulations using the POWHEG event generator~\cite{powheg}.

  \subsection{MC simulations}
  \label{sec:simu}

The production of muons from \W and \Z/$\gamma^*$ decays was simulated by means of MC simulations at NLO using the POWHEG event generator~\cite{powheg}. Since POWHEG is only intended for the simulation of hard partonic scattering processes, it was matched to PYTHIA 6~\cite{pythia} for parton shower description. In the simulations, the CT10 PDF set~\cite{ct10} was used along with the EPS09NLO~\cite{eps09} parametrisation of the nuclear modifications. In order to account for the isospin effect, which is of particular importance for the \W-boson production yields, simulations of proton--proton (pp), proton--neutron (pn), neutron--proton (np), and also neutron--neutron (nn) binary collisions for \PbPb, were performed. The total cross sections were obtained from the single pp, pn, np, and nn cross sections combined with weights proportional to the density of protons and neutrons in a Pb nucleus:
\begin{equation}
    \frac{\dd^2 \sigma^{\rm pPb}_{\rm NN}}{\dd \pt \dd y} = \frac{Z}{A} \times \frac{\dd^2 \sigma^{\rm pPb}_{\rm pp}}{\dd \pt \dd y} + \frac{A-Z}{A} \times \frac{\dd^2 \sigma^{\rm pPb}_{\rm pn}}{\dd \pt \dd y},
    \label{eqn:iso_ppb}
\end{equation}
\begin{equation}
    \frac{\dd^2 \sigma^{\rm PbPb}_{\rm NN}}{\dd \pt \dd y} = \frac{Z^2}{A^2} \times \frac{\dd^2 \sigma^{\rm PbPb}_{\rm pp}}{\dd \pt \dd y} + \frac{(A-Z)^2}{A^2} \times \frac{\dd^2 \sigma^{\rm PbPb}_{\rm nn}}{\dd \pt \dd y}
            + \frac{Z(A-Z)}{A^2} \times \left( \frac{\dd^2 \sigma^{\rm PbPb}_{\rm pn}}{\dd \pt \dd y} + \frac{\dd^2 \sigma^{\rm PbPb}_{\rm np}}{\dd \pt \dd y} \right),
    \label{eqn:iso_pbpb}
\end{equation}
where Eq.~\ref{eqn:iso_ppb} indicates the combination in \pPb collisions and Eq.~\ref{eqn:iso_pbpb} the combination for the \PbPb system.

The contribution of muons from heavy-flavour hadron decays was simulated using the Fixed-Order Next-to-Leading-Log (FONLL) approach~\cite{fonll}. The FONLL calculations were performed with the NNPDF3.1 PDF set~\cite{nnpdf}, without accounting for nuclear modifications. In \pPb collisions, the nuclear effects mainly affect the production of heavy-flavour hadrons at low \pt, typically below 5 \GeVc~\cite{alice_hfRpa}, and are expected to be negligible in the \pt interval studied in this paper. In the analysis of the \PbPb data sample, the FONLL predictions were multiplied by the nuclear modification factor $R_{\rm AA}$ of muons from heavy-flavour hadron decays, taken from simulations performed within the EPOS framework~\cite{epos} in the interval $10 < \pt^\mu < 50$ \GeVc, fitted with a first-order polynomial function and further extrapolated to high \pt. The FONLL predictions were then used as inputs for the MC generation of muons from heavy-flavour hadron decays.

The MC simulations were performed by using the GEANT3 transport code~\cite{geant3} combined with a detailed simulation of the detector response and taking into account the time evolution of the detector configuration and alignment effects. In the high-$\pt^\mu$ region studied in this analysis ($\pt^\mu > 10$ \GeVc), the tracks are weakly bent, the alignment of the tracking chambers is therefore of utmost importance for the track reconstruction. The absolute positions of the chambers were first measured with photogrammetry before the data taking. The relative positions of the detection elements were then refined with a combination of reconstructed tracks in data samples recorded with and without magnetic field using a modified version of the MILLEPEDE package~\cite{millipede}, up to a precision of about 100 $\mu$m. The estimated residual misalignment is then taken into account in the MC simulations. In addition, one may expect a misalignment of the spectrometer in its entirety, which is addressed by studying the track-to-cluster residual distribution in the data and the simulation. The simulation of the tracking chamber response relies on a data-driven parametrisation of the measured resolution of the clusters associated to a track. The distribution of the difference between the cluster and the track positions in each chamber is described using extended Crystal Ball (CB)~\cite{crystal_ball} functions, with parameters tuned on data. The CB parametrisation is then used to reproduce the smearing of the track parameters in the simulations. A global misalignment of the detector is mimicked by shifting the distribution of the track deviations in the magnetic field. The sign of the shift is reverted for positive and negative tracks, and according to the magnetic field polarity. Its magnitude was tuned in order to reproduce the observed difference in the $\pt^\mu$ distribution of positive and negative tracks.

  \subsection{Signal extraction and efficiency correction}
  \label{subsec:signal_extraction}

\begin{figure}[tb]
    \begin{center}
    \includegraphics[width = 0.49\textwidth]{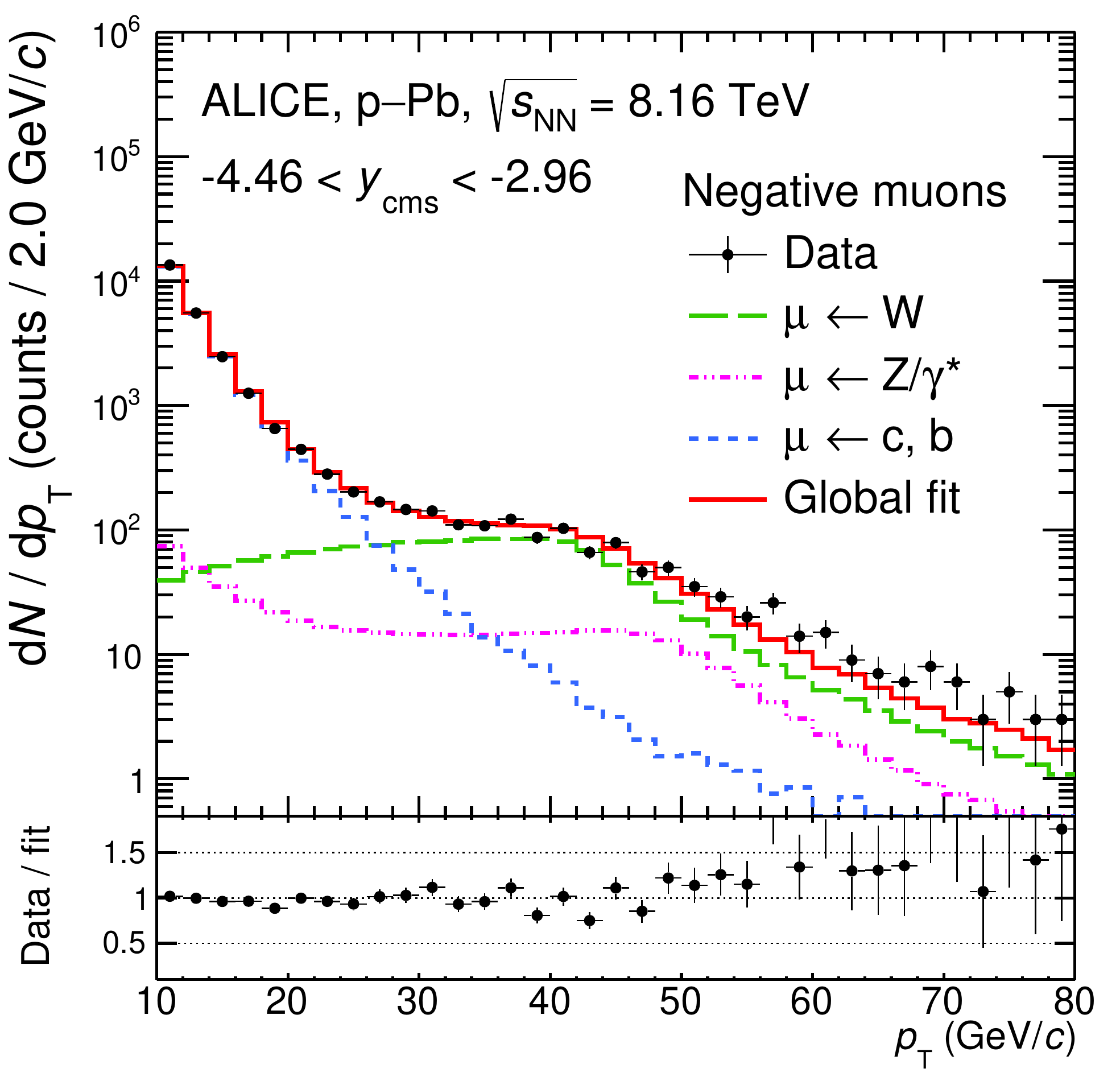}
    \includegraphics[width = 0.49\textwidth]{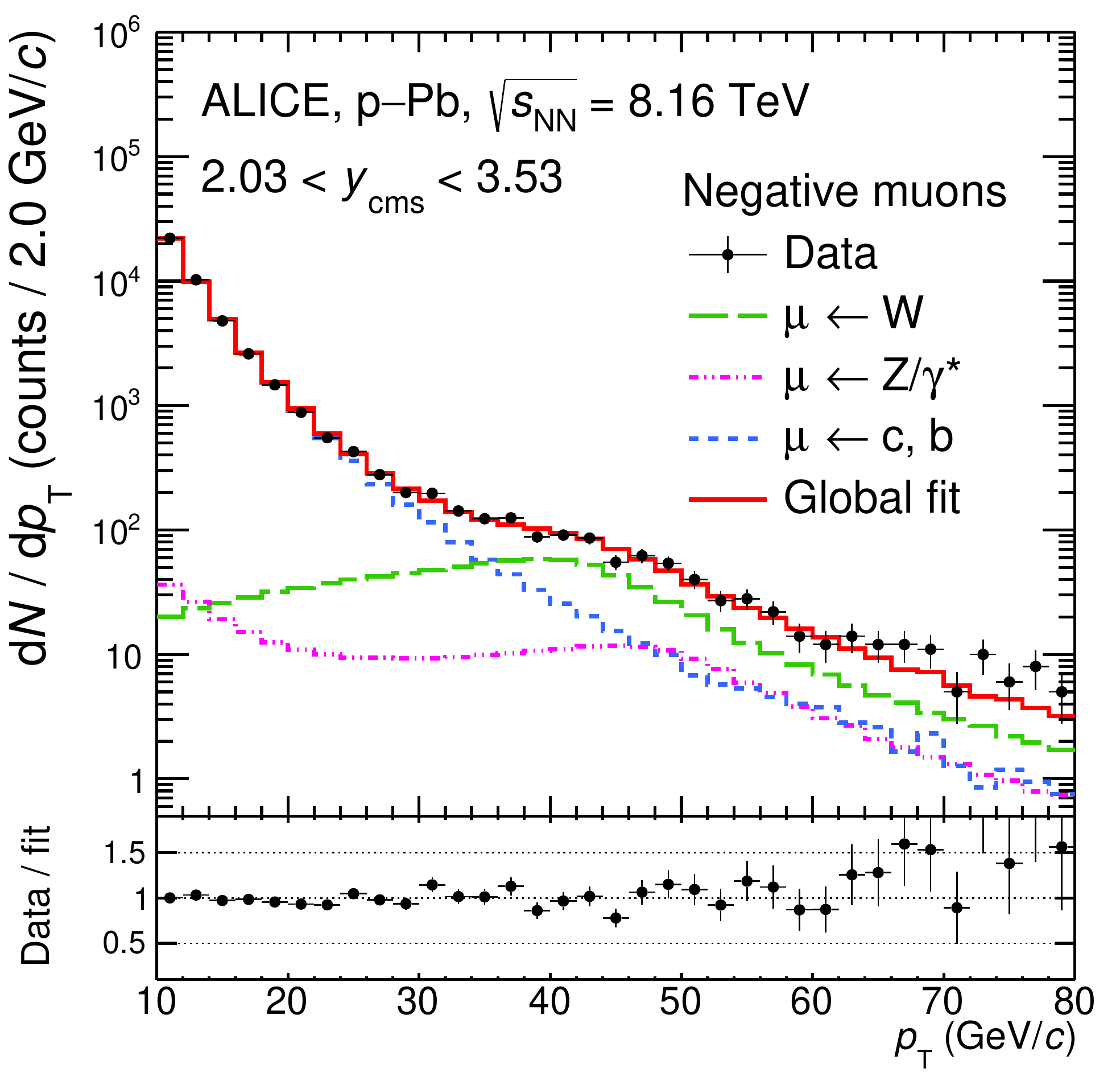}
    \includegraphics[width = 0.49\textwidth]{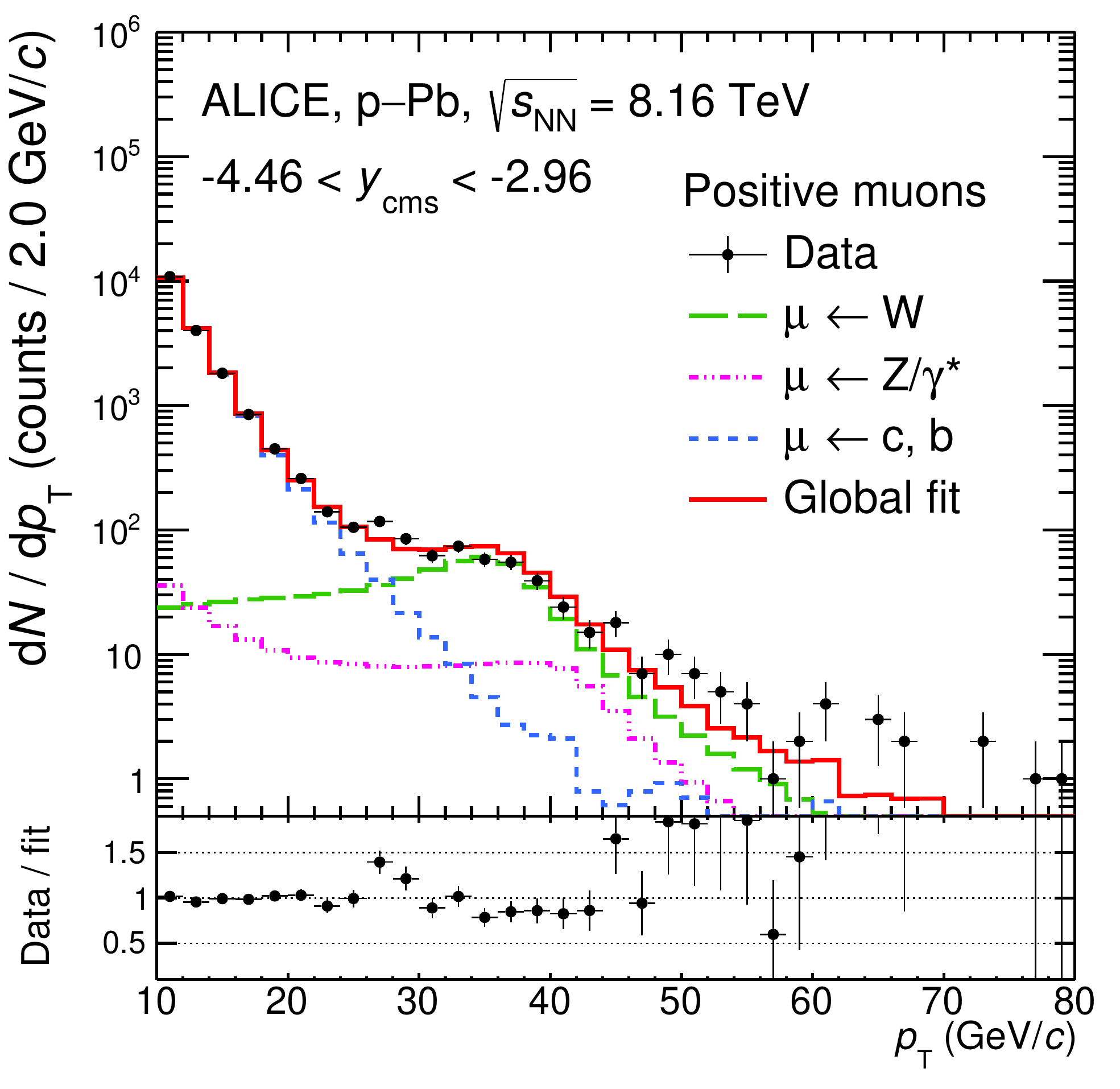}
    \includegraphics[width = 0.49\textwidth]{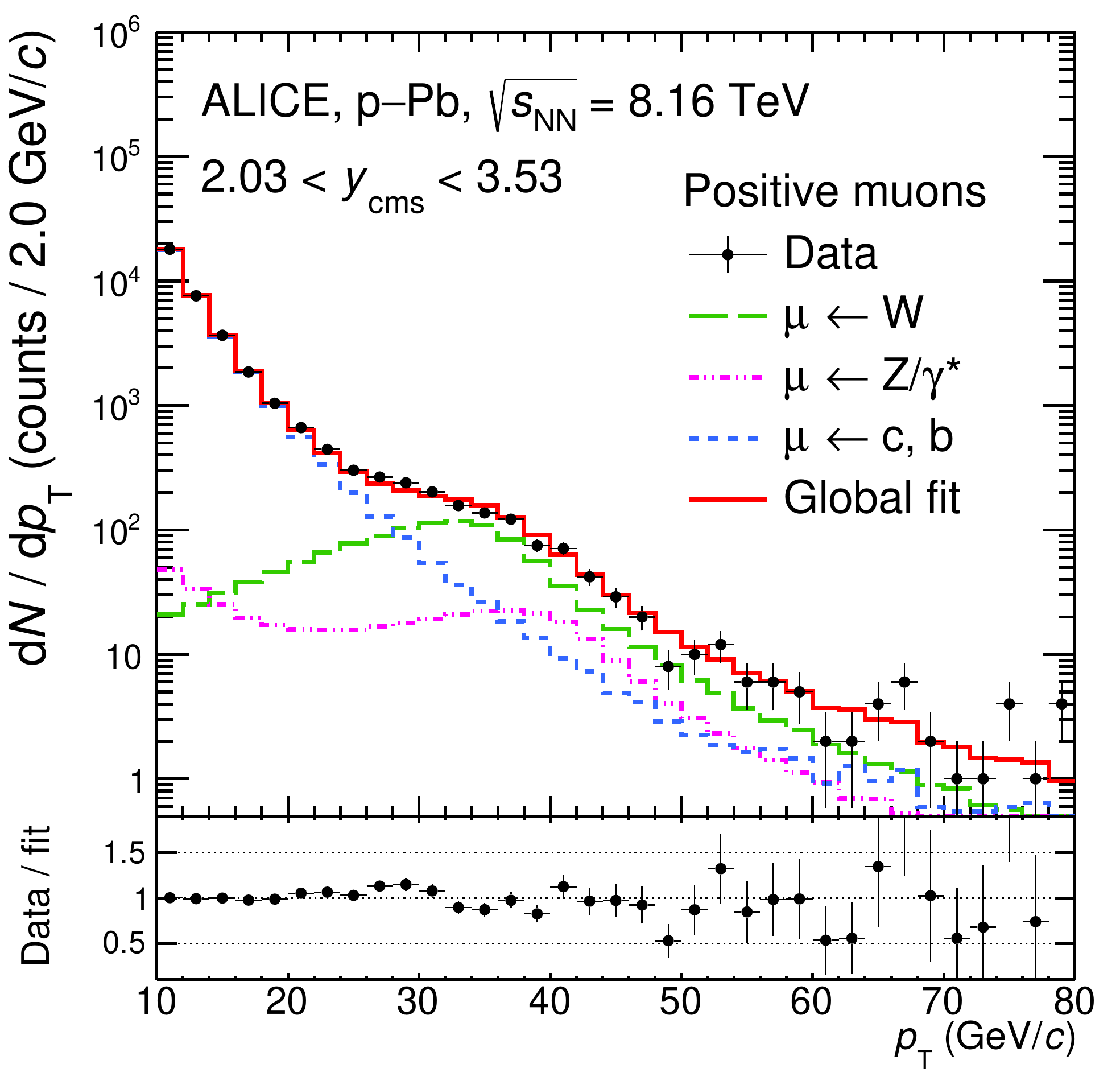}
    \end{center}
    \caption{Inclusive transverse momentum distribution of negative (top) and positive (bottom) muons at backward (left) and forward (right) rapidity in \pPb collisions at \eightnn. The results of the fit to the inclusive spectrum using a combination of MC templates is shown as a continuous line, the green, pink and blue dashed lines representing the contributions of the \W-, \Z/$\gamma^*$- and heavy-flavour hadron decay muons, respectively. The bottom panels show the ratio of the data to the fit result.}
    \label{fig:pPb_signal_extraction}
\end{figure}

\begin{figure}[tb]
    \begin{center}
    \includegraphics[width = 0.49\textwidth]{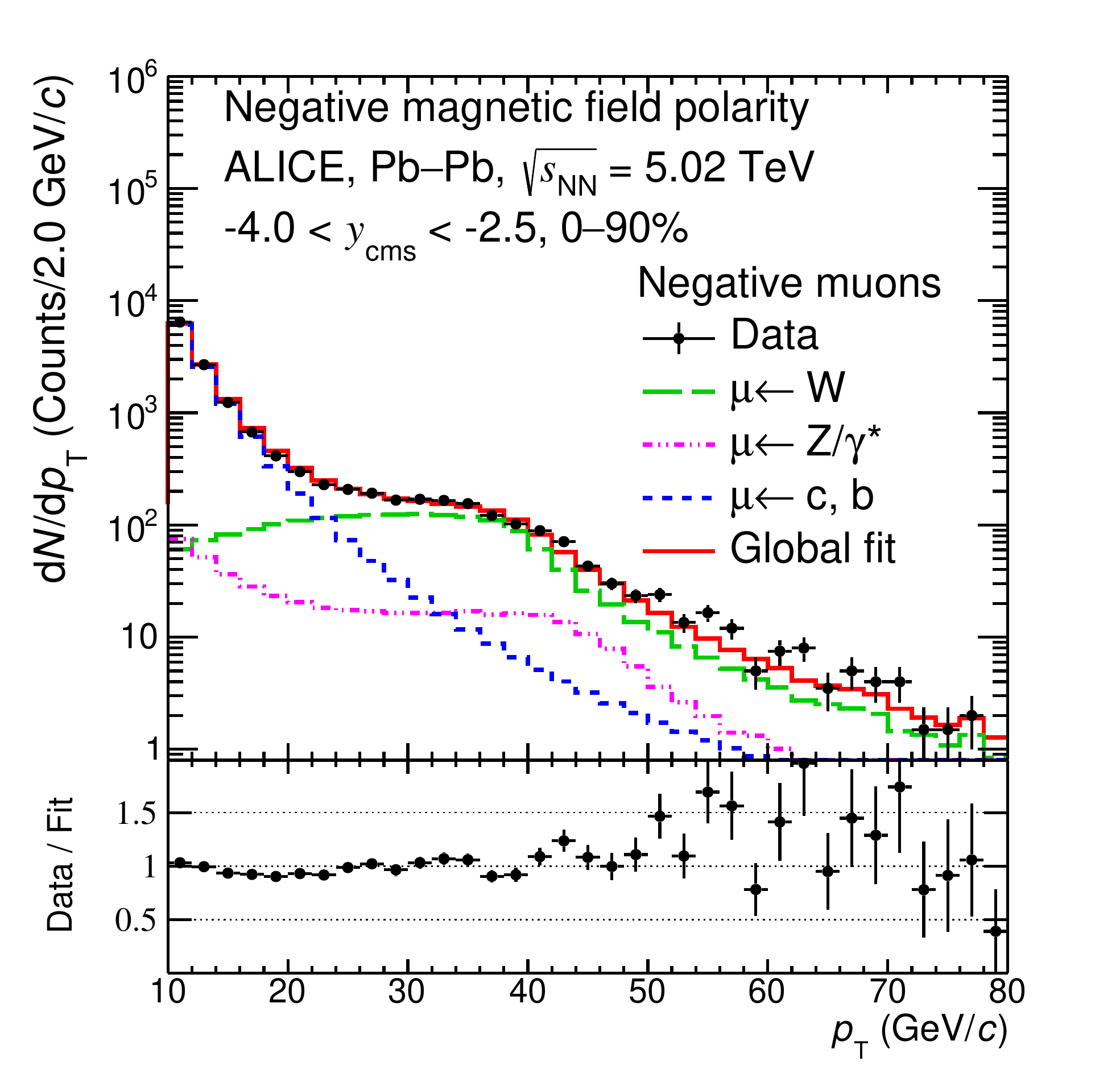}
    \includegraphics[width = 0.49\textwidth]{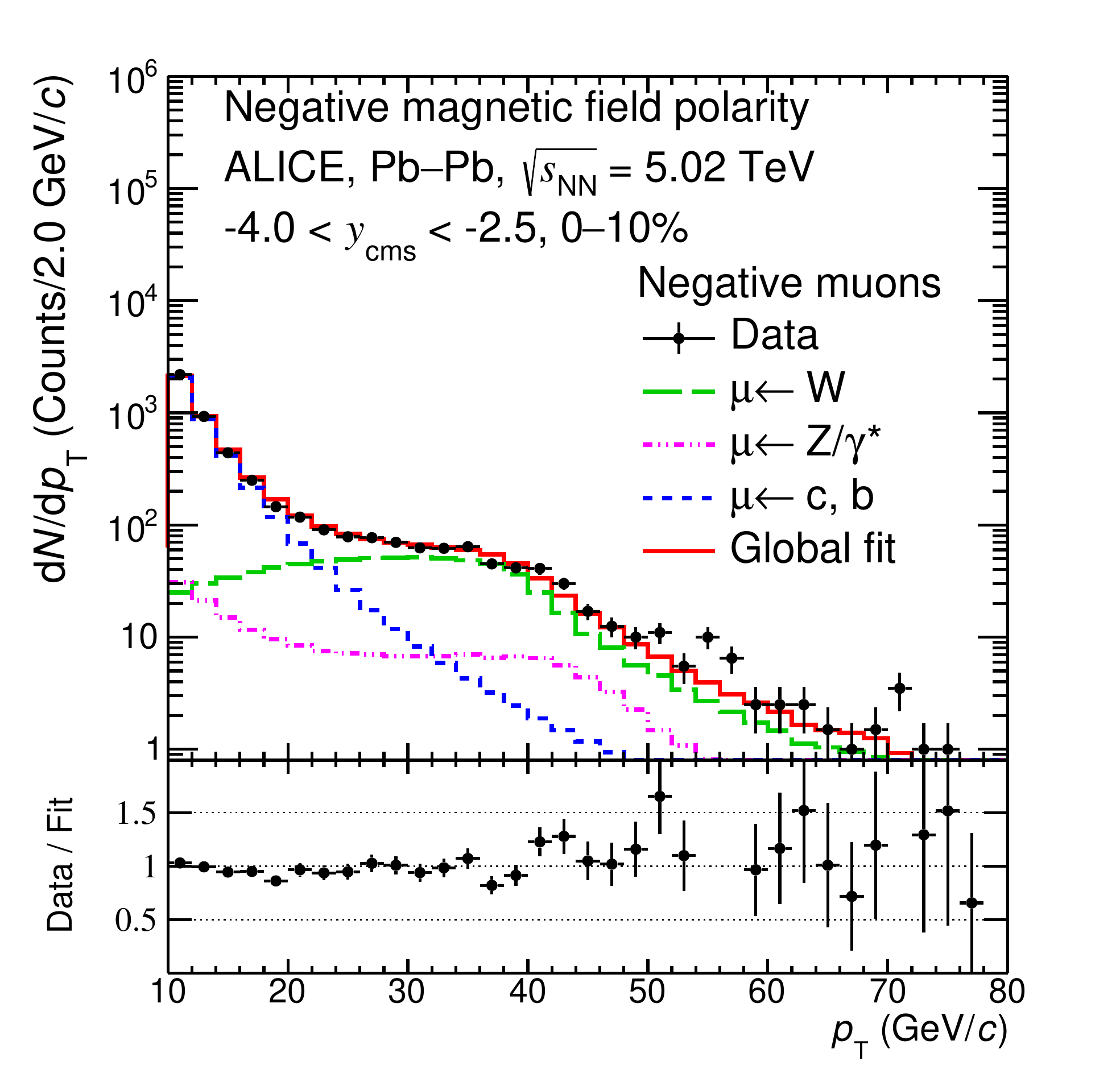}\\
    \includegraphics[width = 0.49\textwidth]{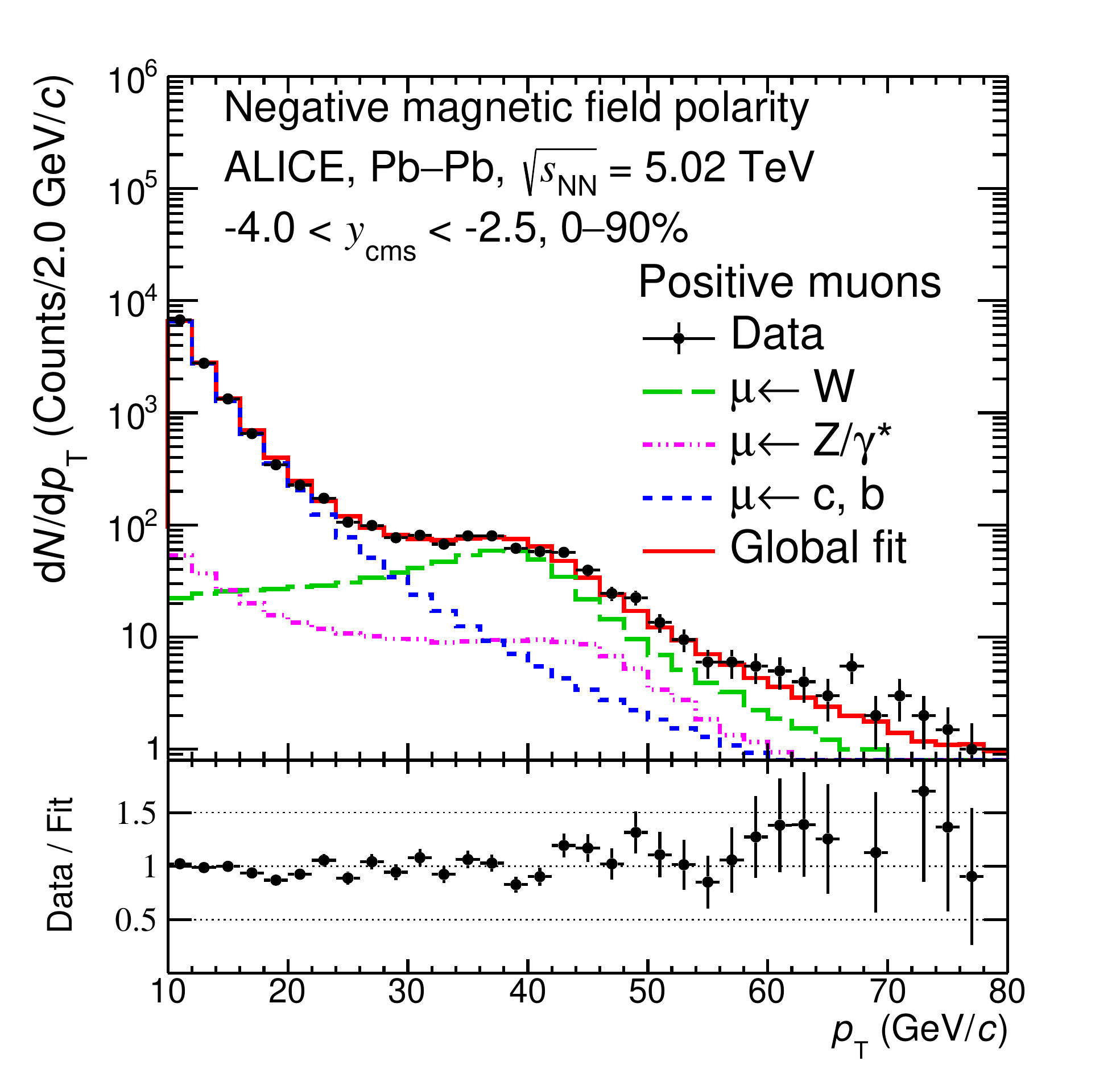}
    \includegraphics[width = 0.49\textwidth]{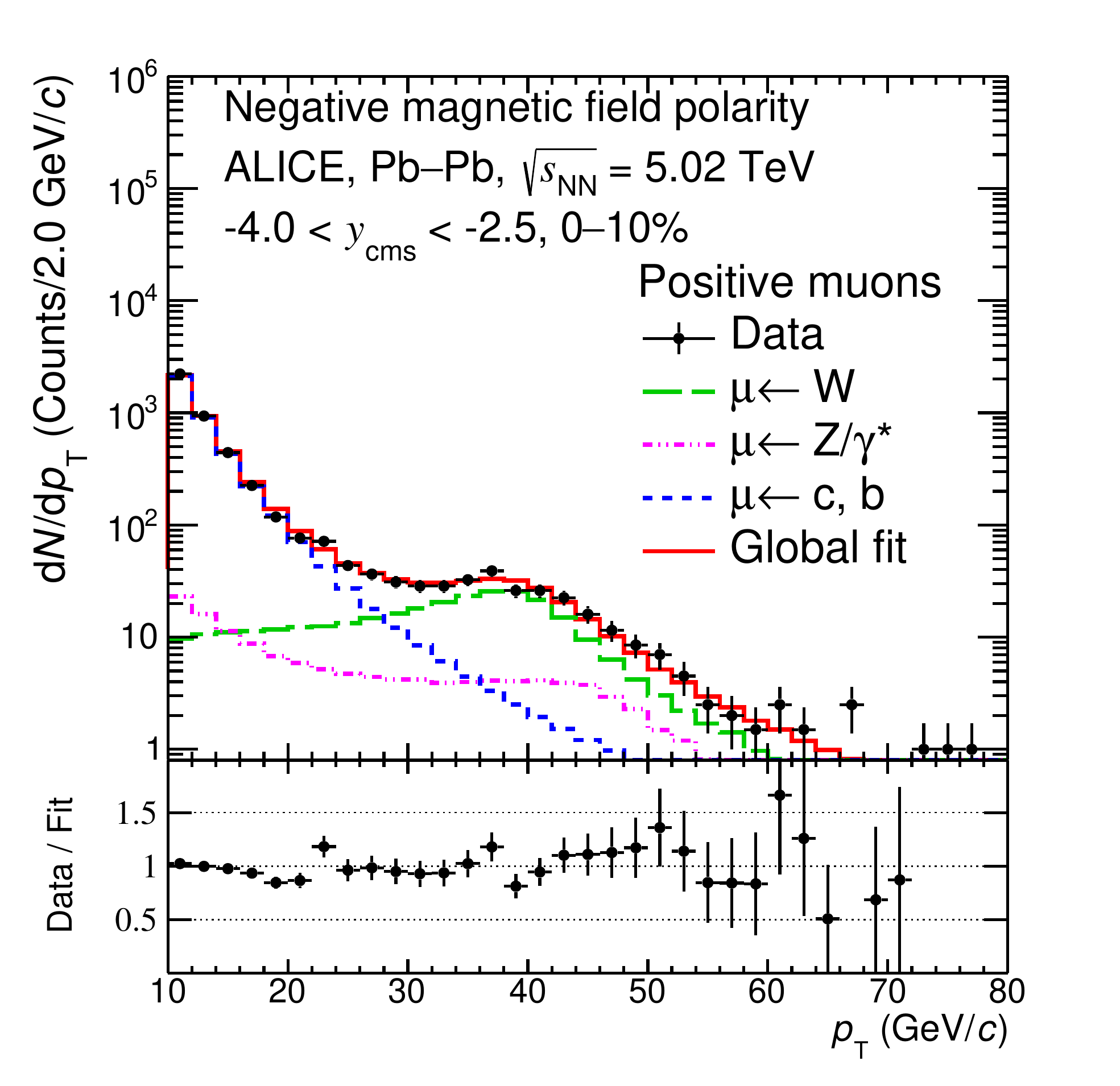}
    \end{center}
    \caption{Inclusive transverse momentum distribution of negative (top) and positive (bottom) muons for the 0--90\% (left) and 0--10\% (right) centrality intervals in \PbPb collisions at \fivenn. The results of the fit to the inclusive spectrum using a combination of MC templates is shown as a continuous line, the green, pink and blue dashed lines representing the contributions of the \W-, \Z/$\gamma^*$- and heavy-flavour hadron decay muons, respectively. The bottom panels show the ratio of the data to the fit result.}
    \label{fig:PbPb_signal_extraction}
\end{figure}

Examples of the \Wminus and \Wplus signal extraction are shown in Fig.~\ref{fig:pPb_signal_extraction} and~\ref{fig:PbPb_signal_extraction} for \pPb and \PbPb collisions, respectively. In \pPb collisions, an example is given for each combination of the colliding beam configuration and the charge of the muon. In \PbPb collisions, examples are given for the two charges of the muon, in the full centrality interval or for the 10\% most central collisions. For both collision systems, the decay of \W bosons becomes the dominant contribution for $\pt^\mu$ above 25 or 30 \GeVc. The fits to Eq.~\ref{eqn:signal_extraction} are found to describe well the data, although at high $\pt^\mu$ they tend to underestimate the muon yield in some configurations. This difference between the data and the fit occurs in a $\pt^\mu$ interval where the number of muons is small, and has a negligible impact on the signal extraction.

The signal extraction procedure is affected by different sources of systematic uncertainties, which are related to the knowledge of the shape of the templates. The effect of this uncertainty on the extracted \W-boson yield was estimated by studying the fit stability with reasonable variations of these shapes. The \W-boson and \Z/$\gamma^*$ templates were generated using the CT10~\cite{ct10} and CTEQ6~\cite{cteq6} PDF sets paired with either EPS09~\cite{eps09} or EKS98~\cite{eks98} nPDF, both at either LO or NLO. Varying the inputs of the simulations leads to different values of the $R$ factor of Eq.~\ref{eqn:signal_extraction}, estimated from the same simulations. The template accounting for muons from heavy-flavour hadron decays was computed by varying the FONLL calculations used as input within their uncertainties, originating from the choice of quark masses, factorisation and renormalisation scales, and from the uncertainty on the PDFs. In \PbPb collisions the uncertainty due to the $\pt^\mu$ extrapolation of the $R_{\rm AA}$ of muons from heavy-flavour hadron decays was estimated using different functional forms, as well as fitting the ALICE measurement of the $R_{\rm AA}$ between $7 < \pt^\mu < 20$ \GeVc and extrapolating the fit result to high \pt. The difference between the various extrapolations is taken as systematic uncertainty on the FONLL weighting procedure. For the simulation of the detector response, the tuning parameter of the global shift was varied within the uncertainty on its determination. The CB parameters for the cluster resolution, obtained from the data-driven method, were replaced by a set of parameters evaluated from simulations. The fit range was varied by moving the lower limit of the $\pt^\mu$ interval between 10 and 20 \GeVc and the higher limit between 50 and 80 \GeVc. All the possible combinations of the variations were considered, each configuration yielding a value for $N^{\rm raw}_{\muonpm \leftarrow \W}$. The combined $\chi^2/$ndf of the fits to the \muon and \Amuon distributions was required to be smaller than 2 to ensure that only the configurations able to satisfactorily reproduce the data were kept. The final number of muons from \W decays, and the associated statistical uncertainty, were obtained by averaging over the $N^{\rm raw}_{\muonpm \leftarrow \W}$ distribution obtained from all considered variations.

The extracted raw yield is corrected for the detection and reconstruction efficiency $\epsilon$ obtained from the simulations described in the previous section. The efficiency is estimated as the ratio of the number of reconstructed muons from \W-boson decays, with the same selections as applied to the data, to the number of generated \W-decay muons in the region of interest, that is the fiducial region defined by the selection on the muon $\pt^\mu >$ 10 \GeVc, and the detector angular acceptance, $2.5 < y^\mu_{\rm cms} < 4$. The efficiency in \pPb collisions amounts to 90\% (91\%) in the p-going configuration and 88\% (89\%) in the Pb-going one for \muon (\Amuon). In \PbPb collisions, the efficiency is additionally affected by the detector occupancy. This effect was taken into account by embedding the simulated signal into \PbPb data. The efficiency for the most central collisions is found to be 94\% of the efficiency of the most peripheral collisions. The centrality-integrated efficiency for the 2015 period amounts to 83\% and 81\% for \muon and \Amuon, respectively, while for the 2018 period the efficiency is 80\% and 79\% for \muon and \Amuon, respectively. The efficiency has no significant dependence on $\pt^\mu$, and decreases by about 9\% from the most central to the largest rapidities.

  \subsection{Systematic uncertainties}

The systematic uncertainties are summarised in Table~\ref{table:syst}. The signal extraction procedure described in the previous section yields a distribution of $N^{\rm raw}_{\muonpm \leftarrow \W}$ after the variation of the fit configuration and the simulation parameters. The dispersion (RMS) of the distribution was used as systematic uncertainty on the signal extraction. The uncertainty originating from the signal extraction procedure ranges from about 4\% to 9\% in the rapidity- and centrality-integrated studies. In the rapidity-differential measurements, for the largest rapidity intervals, the lower amount of signal reduces the stability of the fit such that the systematic uncertainty rises up to 22\%.

\begin{table}
  \centering
  \caption{Summary of systematic uncertainties affecting the \W-boson measurements in \pPb and \PbPb collisions. The values given for \PbPb collisions are for the combined 2015 and 2018 data samples. The ranges correspond to the largest variations found in differential analyses.}
  \begin{tabular}{|c|c c|c|}
    \hline
    \multirow{2}{*}{Source}       & \multicolumn{3}{c|}{Relative systematic uncertainty} \\
                                  \cline{2-4}
                                  & \pPb analysis   & \Pbp analysis    & \PbPb analysis \\
    \hline \hline
    Signal extraction             & 5.9 -- 8.8 \%   & 3.8 -- 7.3 \%    & 2.9 -- 3.3 \% \\
    - as a function of rapidity   & 3.9 -- 14.3 \%  & 2.5 -- 22 \%     & --- \\
    - as a function of centrality & 5.1 -- 9.7 \%   & 3.6 -- 9.0 \%    & 3.0 -- 7.4 \% \\
    \hline
    Tracking efficiency           & 0.5 \%          & 1.0 \%           & 1.5 \% \\
    Trigger efficiency            & \multicolumn{2}{c|}{0.5 \%}        & 0.75 \% \\
    Trigger--tracker matching     & \multicolumn{2}{c|}{0.5 \%}        & 0.5 \% \\
    Alignment                     & \multicolumn{2}{c|}{0.1 -- 1.2 \%} & 1.8 \% \\
    \hline
    Normalisation factor          & 1.4 \%          & 1.1 \%           & 1.0 \% \\
    $\sigma_{\rm V0M}$            & \multicolumn{2}{c|}{1.9 \%}        & 2.0 \% \\
    \hline
    \avNcollMult                  & \multicolumn{2}{c|}{2.8 -- 4.3 \%} & --- \\
    \avTaa                        & \multicolumn{2}{c|}{---}           & 0.7 -- 2.0 \% \\
    \hline
  \end{tabular}
  \label{table:syst}
\end{table}

The uncertainty of the efficiency computation is evaluated by varying the simulation environment. It was observed that, in the simulations, only the ability to properly reproduce the alignment conditions provides a significant source of uncertainty through the estimation of the CB tails parameters and the tuning of the parameter accounting for the global shift. The systematic uncertainty is taken as the largest difference between the efficiencies computed with all the possible configurations. The uncertainty on the tracking efficiency is obtained by considering the difference between the efficiencies obtained from data and MC simulations, using the redundancy of the tracking chamber information~\cite{alice_perf3}. The uncertainty on the muon trigger efficiency is determined by propagating the uncertainty on the intrinsic efficiency of the individual trigger chambers, which is evaluated using a data-driven method based on the redundancy of the trigger chamber information~\cite{alice_perf3}. The choice for the $\chi^2$ value in defining the  matching between the tracks in the tracking and trigger systems introduces an additional 0.5\% uncertainty. The difference between the two methods for the computation of the normalisation, detailed in Section~\ref{sec:selection}, is taken as its systematic uncertainty. The uncertainties on the $\sigma_{\rm V0M}$ values are taken from Refs.~\cite{lumi-pPb8tev,lumi-PbPb5tev} where their evaluation is detailed. Finally, the uncertainty on \avNcollMult in \pPb collisions is evaluated as the difference with respect to the average number of binary collisions estimated using an alternative method based on the multiplicity measured in the Pb-going direction~\cite{centrality}. In \PbPb collisions, the uncertainty on \avTaa is estimated by varying the parameters of the Glauber model within their own uncertainties, adding in quadrature the maximum-to-average ratio of the upward and downward variations from all sources~\cite{centrality}. The total systematic uncertainty is obtained by summing all the considered sources in quadrature.

\section{Results}
\label{sec:results}

  \subsection{\pPb collisions}
  \label{sec:pPb}

    \subsubsection{Production cross sections}

The $\muonpm \leftarrow \W$ rapidity-differential production cross section, uncorrected for the W-to-muon branching ratio BR, is evaluated as
\begin{equation}
  \frac{\dd \sigma_{\W \rightarrow \muonpm \num}}{\dd y} = \frac{N_{\muonpm \leftarrow \W}}{\Delta y \times \epsilon \times \lumi_{\rm int}},
  \label{eqn:xSec}
\end{equation}
where $N_{\muonpm \leftarrow \W}$ is the measured yield of muons from \W decays, $\Delta y$ is the width of the rapidity interval, $\epsilon$ is the efficiency correction factor, and $\lumi_{\rm int}$ the integrated luminosity. In \pPb collisions at \eightnn, the values of the corresponding production cross sections are reported in Table~\ref{table:xSec-pPb}, where the Pb-going denomination refers to the backward rapidity interval $-4.46 < y^\mu_{\rm cms} < -2.96$ and the p-going denomination to the forward interval $2.03 < y^\mu_{\rm cms} < 3.53$.

\begin{table}[h]
  \centering
  \caption{Rapidity-differential production cross sections of \Wminus and \Wplus bosons measured from their muonic decays in \pPb collisions at \eightnn, for muons with $\pt^\mu > 10$ \GeVc.}
  \renewcommand{\arraystretch}{1.5}
  \begin{tabular}{|c|c|c|}
    \cline{2-3}
    \multicolumn{1}{c|}{}               & $-4.46 < y^\mu_{\rm cms} < -2.96$ (Pb-going)                         & $2.03 < y^\mu_{\rm cms} < 3.53$ (p-going) \\
    \hline
    $\Wminus \rightarrow \muon \Anum$ & $105.4 \pm 3.7 \text{ (stat)} \pm 5.2 \text{ (syst)} \text{ nb}$ & $90.2 \pm 4.8 \text{ (stat)} \pm 8.2 \text{ (syst)} \text{ nb}$ \\
    $\Wplus \rightarrow \Amuon \num$  & $37.1 \pm 2.1 \text{ (stat)} \pm 2.9 \text{ (syst)} \text{ nb}$  & $120.8 \pm 5.2 \text{ (stat)} \pm 7.7 \text{ (syst)} \text{ nb}$ \\
    \hline
  \end{tabular}
  \label{table:xSec-pPb}
\end{table}

\begin{figure}
    \begin{center}
    \includegraphics[width = 0.95\textwidth]{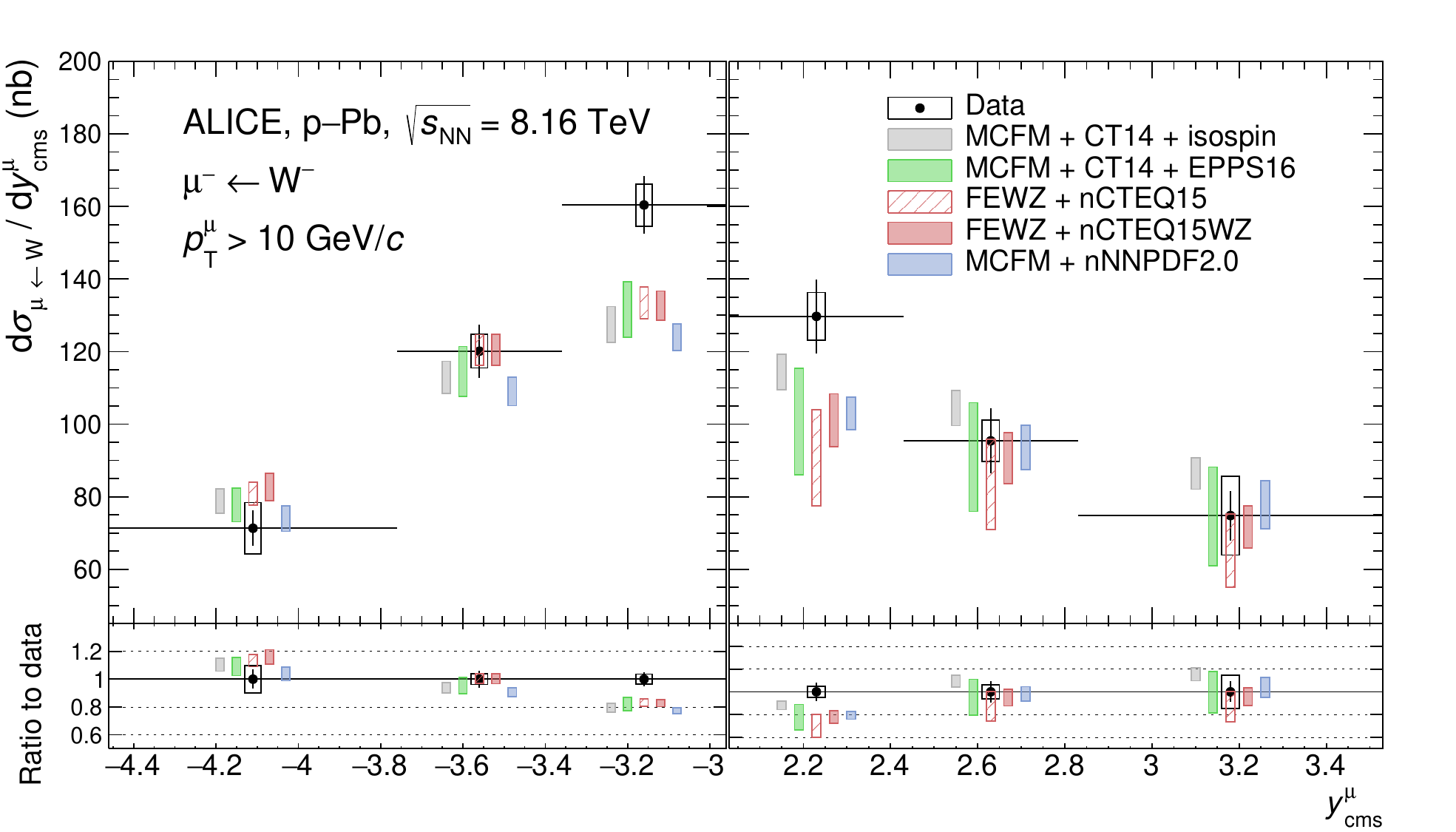}
    \includegraphics[width = 0.95\textwidth]{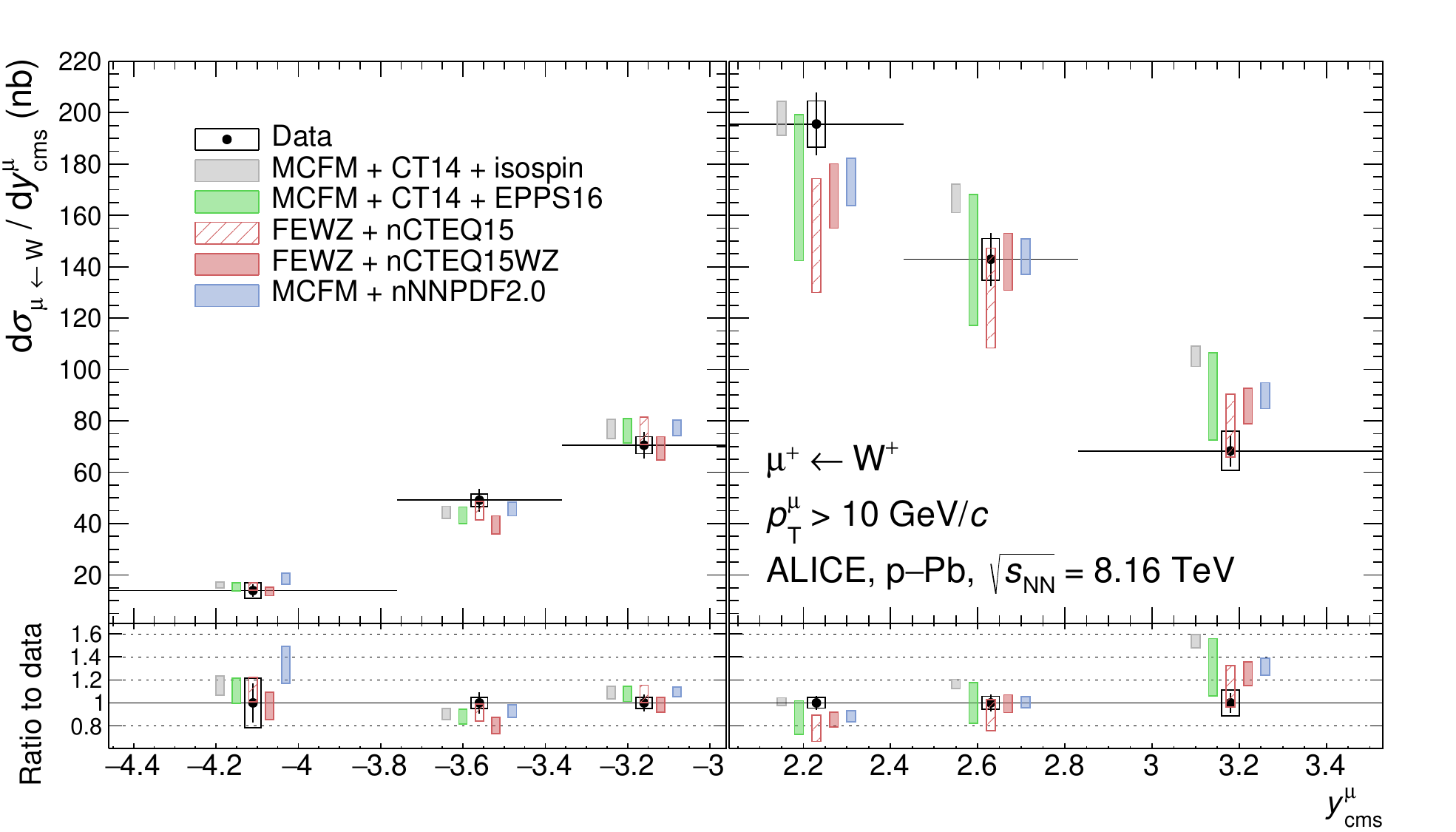}
    \end{center}
    \caption{Production cross section of muons from \Wminus (top) and \Wplus (bottom) decays as a function of rapidity for muons with $\pt^\mu > 10$ \GeVc in \pPb collisions at \eightnn. The measurements are compared with predictions from several nPDF sets, as well as with calculations based on the CT14 PDF set~\cite{ct14} without nuclear modifications of the PDF. All the calculations include the isospin effect. The bottom panels show the ratio of the calculations to the measured production cross section. The horizontal bars correspond to the width of the rapidity intervals. The vertical bars and boxes indicate the statistical and systematic uncertainties, respectively. The data points are placed at the centres of the rapidity intervals, while the theory predictions are horizontally shifted for better visibility.}
    \label{fig:pPb_xSec}
\end{figure}

The production cross section is shown as a function of rapidity, in the Pb-going and p-going directions and for both charges of the W boson, in Fig.~\ref{fig:pPb_xSec}. The measurements are compared with several pQCD calculations, based on Monte Carlo for FeMtobarn processes (MCFM)~\cite{mcfm} or Fully Exclusive W and Z production (FEWZ)~\cite{fewz} simulations. The MCFM and FEWZ codes enable the calculation of hard processes in hadronic collisions, involving heavy flavour and top quarks, electroweak bosons and the Higgs boson. The two codes were shown to produce similar predictions of the electroweak-boson production at NLO~\cite{alekhin2021}. The nuclear modifications are computed using the CT14+EPPS16~\cite{epps16}, nCTEQ15WZ~\cite{ncteq15wz} and nNNPDF2.0~\cite{nnnpdf} parametrisations, as discussed in Section~\ref{sec:intro}. To illustrate the effect of using the LHC data in the determination of nPDFs, predictions were also obtained from the nCTEQ15 set~\cite{ncteq15} in which no LHC data were included. In order to disentangle the effect of the nuclear modifications of the PDFs from other effects affecting the \W-boson production, such as the isospin, predictions are shown for the CT14 PDF~\cite{ct14} without nuclear modifications. All calculations are performed at NLO, the proton and neutron contributions are weighted following the nucleon content of the Pb ion to reproduce the isospin dependence of the \W-boson production.

Several effects affect the production of the \Wminus and \Wplus bosons in \pPb collisions. The isospin effect, originating from the difference in the quark content of the Pb nucleus to that of the proton, increases the production of \Wminus and decreases that of \Wplus. The rapidity shift due to the asymmetric system pushes the forward rapidity range covered by the muon spectrometer, corresponding to the p-going configuration, towards midrapidity, where the production cross section is higher, and moves the backward rapidity range, in the Pb-going configuration, towards even larger rapidities where the production rate is reduced. Moreover, the production is affected by the helicity conservation. The weak interaction only couples left-handed fermions and right-handed antifermions. For angular momentum conservation, the outgoing fermion \muon (antifermion \Amuon) follows the direction of the incoming quark q (antiquark $\overline{\mathrm{q}}$). The production cross section is then maximum when the outgoing lepton (antilepton) goes in the direction of the incoming quark (antiquark). As a result, \Wminus bosons produced at large absolute rapidities will preferably emit \muon in their momentum direction and \Wplus will preferably emit \Amuon in the opposite direction. In the latter case, the muon reaches the large rapidity covered by the spectrometer only if the boson is produced in the opposite direction, at even larger rapidities where the production quickly drops. Finally, the nuclear modifications of the PDFs affect the production at backward and forward rapidities differently. At backward rapidity, the Bjorken-$x$ interval accessible with the ALICE measurements is influenced by the anti-shadowing and EMC effects, yielding an enhancement and a reduction of the production, respectively. On the other hand, the forward rapidity interval is fully contained within the Bjorken-$x$ region dominated by shadowing, resulting in a suppression of the parton densities. Although in most cases the effects just discussed tend to cancel each other, at least to some extent, they globally act towards a suppression of the \Wplus production at backward rapidities.

The measured \Wplus production cross section is in fair agreement with the model predictions, whereas some tension appears in the description of the rapidity dependence of the \Wminus production cross section, for small values of the absolute rapidity. For \Wplus bosons measured at forward rapidities, corresponding to the shadowing region at low Bjorken-$x$, the measurement favours predictions including the nuclear modifications of the PDFs. The discrepancy with the free-nucleon PDF calculation is especially visible at large positive rapidities where the deviation from the CT14-only prediction reaches 3.5$\sigma$, with the statistical and systematic uncertainties combined quadratically. The precision of the measurement is better than that of the theory, highlighting its ability to provide further constraints for nPDF sets. The comparison between the nCTEQ15 and nCTEQ15WZ predictions shows the impact of the LHC data on the determination of the nPDFs, whose uncertainties are substantially reduced despite the addition of three new free parameters in nCTEQ15WZ, corresponding to the parametrisation of the strange-quark nPDF. The nNNPDF group, which has adopted a methodology based on machine learning for the determination of the nPDF, yields predictions with the smallest uncertainties in the forward rapidity region, corresponding to very low Bjorken-$x$ values. The four models including nuclear modifications are in good agreement with each other, although some discrepancies are present between nCTEQ15WZ and nNNPDF2.0 calculations at backward rapidities. \\

\begin{figure}
    \begin{center}
    \includegraphics[width = 0.95\textwidth]{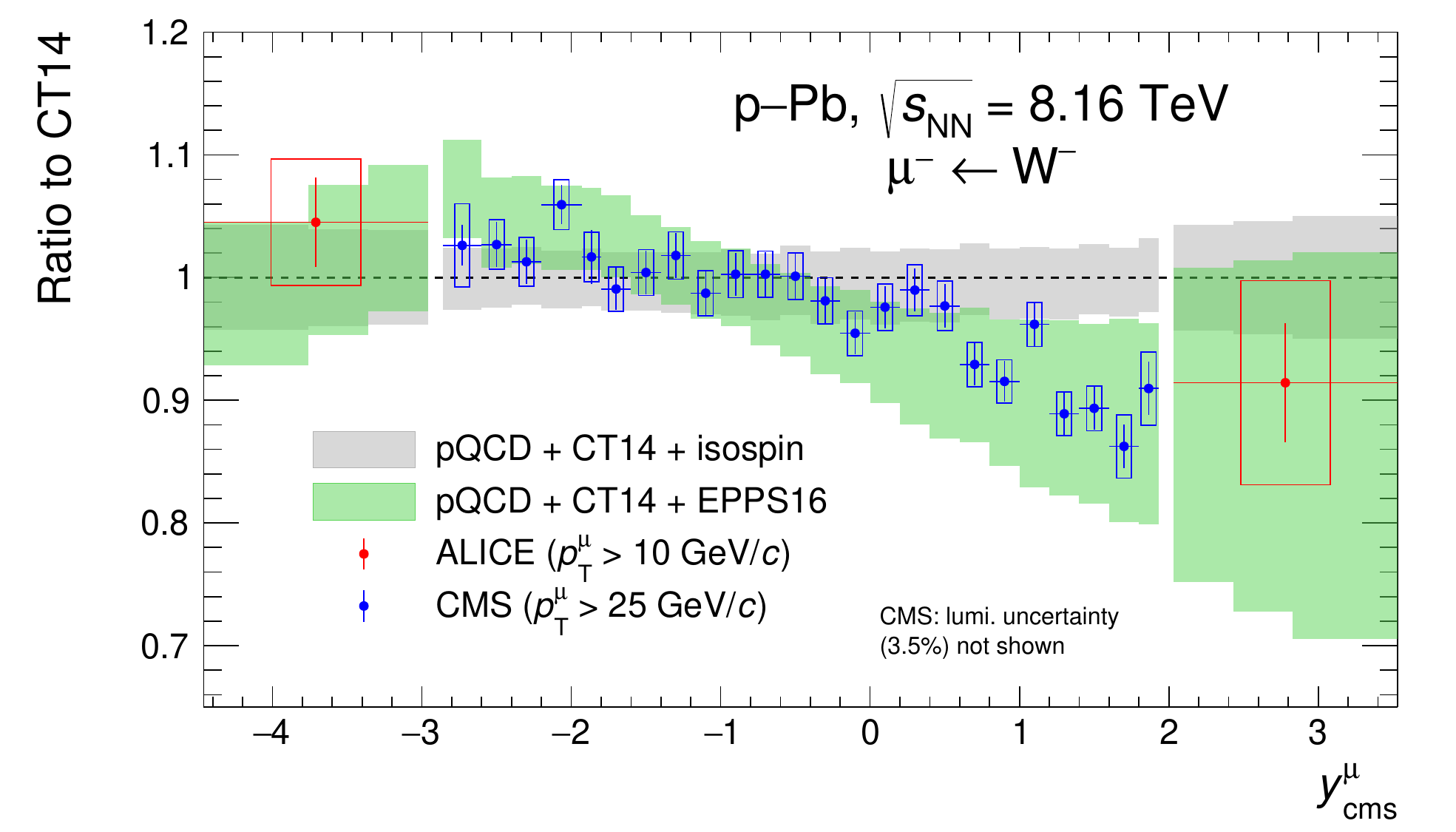}
    \includegraphics[width = 0.95\textwidth]{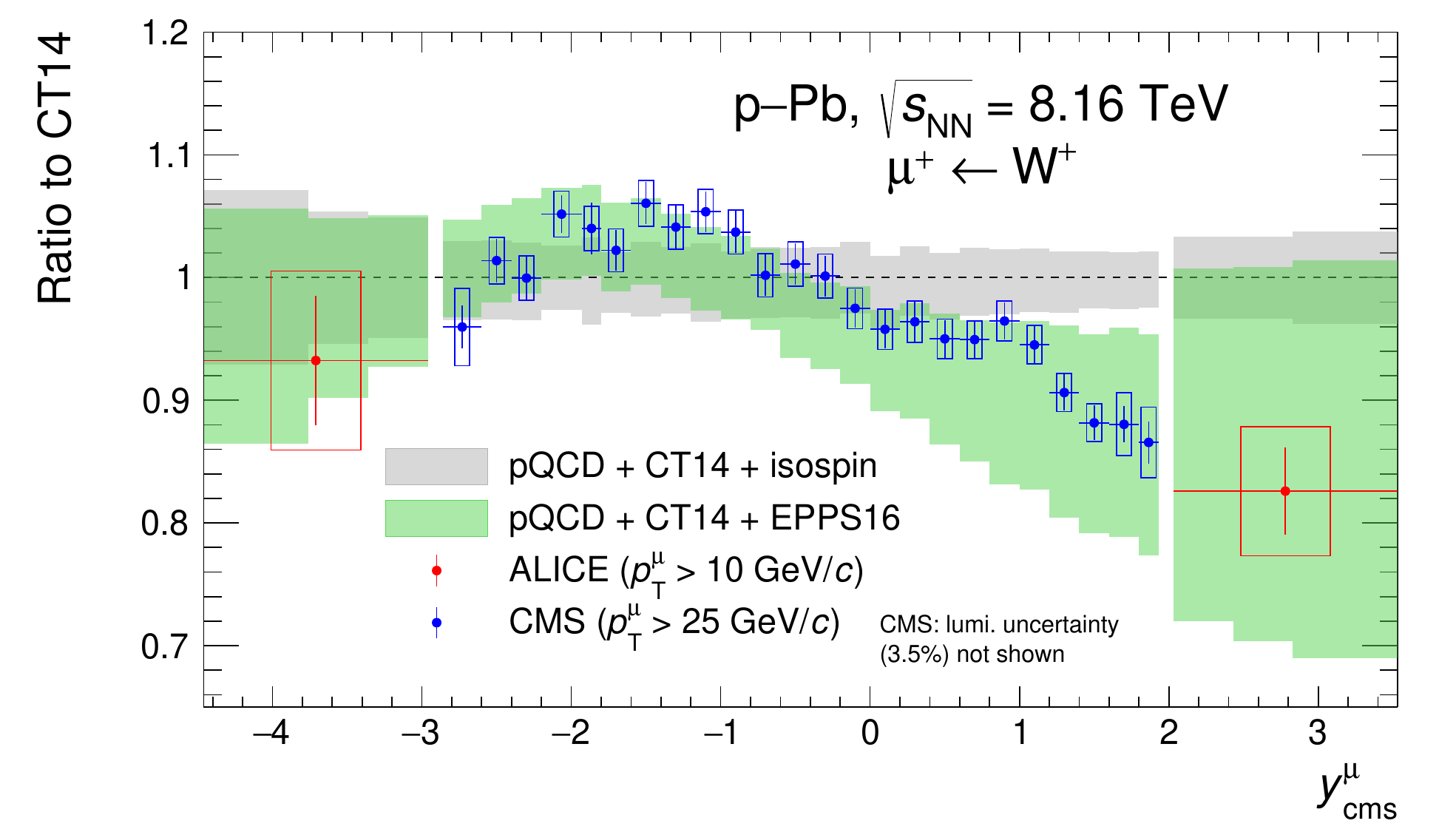}
    \end{center}
    \caption{Ratio to CT14~\cite{ct14} predictions of the production of muons from \Wminus (top) and \Wplus (bottom) decays measured in \pPb collisions at \eightnn by the ALICE and CMS~\cite{Cms:WpPb8tev} Collaborations. The measured ratio is compared to the one obtained from pQCD calculations with CT14+EPPS16~\cite{epps16}. All the calculations include the isospin effect. The grey band around the line at unity indicates the uncertainty on the calculations with CT14 PDFs.}
    \label{fig:pPb_compCMS}
\end{figure}

The CMS Collaboration also measured the production of the \W bosons via the muonic decay channel in \pPb collisions at \eightnn from a data sample with an integrated luminosity of $173.4~\pm~6.1$~nb$^{-1}$~\cite{Cms:WpPb8tev}. The production was measured at midrapidity, in the interval $|\eta^\mu_{\rm lab}| < 2.4$, complementary to the ALICE measurement at large rapidities. A stronger selection was applied on the muon transverse momentum at $\pt^\mu > 25$ \GeVc, a direct comparison is therefore not possible. However, the two measurements can be compared through their agreement with theoretical calculations. Figure~\ref{fig:pPb_compCMS} shows the ratio of the measurements to pQCD calculations performed including the isospin effect and using either the CT14 PDF set (without nuclear modifications) or the CT14 set with the EPPS16 nPDFs.

The measurements of ALICE extend to large rapidities the measurements of the CMS Collaboration in the central region, and support the trend observed at the edge of the CMS rapidity acceptance. The calculations including the EPPS16 nPDFs provide a better description of the data over the whole rapidity interval as compared to the predictions with the CT14 PDFs without nuclear effects, especially for the \Wplus boson.

    \subsubsection{Lepton charge asymmetry}

The production of \Wminus and \Wplus bosons is significantly dependent on the light-quark content of the nucleus. The study of the asymmetry in their production therefore provides a sensitive probe of the up and down nPDF as well as the down-to-up ratio in the nucleus. In this regard, the lepton charge asymmetry $A_{\rm ch}$ can be defined as
\begin{equation}
  A_{\rm ch} = \frac{N^{\rm corr}_{\Amuon \leftarrow \Wplus} - N^{\rm corr}_{\muon \leftarrow \Wminus}}{N^{\rm corr}_{\Amuon \leftarrow \Wplus} + N^{\rm corr}_{\muon \leftarrow \Wminus}},
  \label{eqn:Ach}
\end{equation}
where $N^{\rm corr}_{\muon \leftarrow \Wminus}$ and $N^{\rm corr}_{\Amuon \leftarrow \Wplus}$ are the number of muons from \Wminus and \Wplus decays, respectively, extracted from the data and corrected for the detection and reconstruction efficiency. Part of the experimental uncertainties, such as the trigger and tracking efficiencies, cancels in the calculation of the asymmetry. The theoretical precision is also increased, e.g. through the cancellation of the uncertainties due to the pQCD scales. It should be noted that the lepton charge asymmetry might be much more sensitive to the baseline PDF than to its nuclear modifications~\cite{ZWnPDFconstraints}, possibly enabling the study of the free-nucleon PDF in heavy-ion collisions.

\begin{figure}[h!]
    \begin{center}
    \includegraphics[width = 0.90\textwidth]{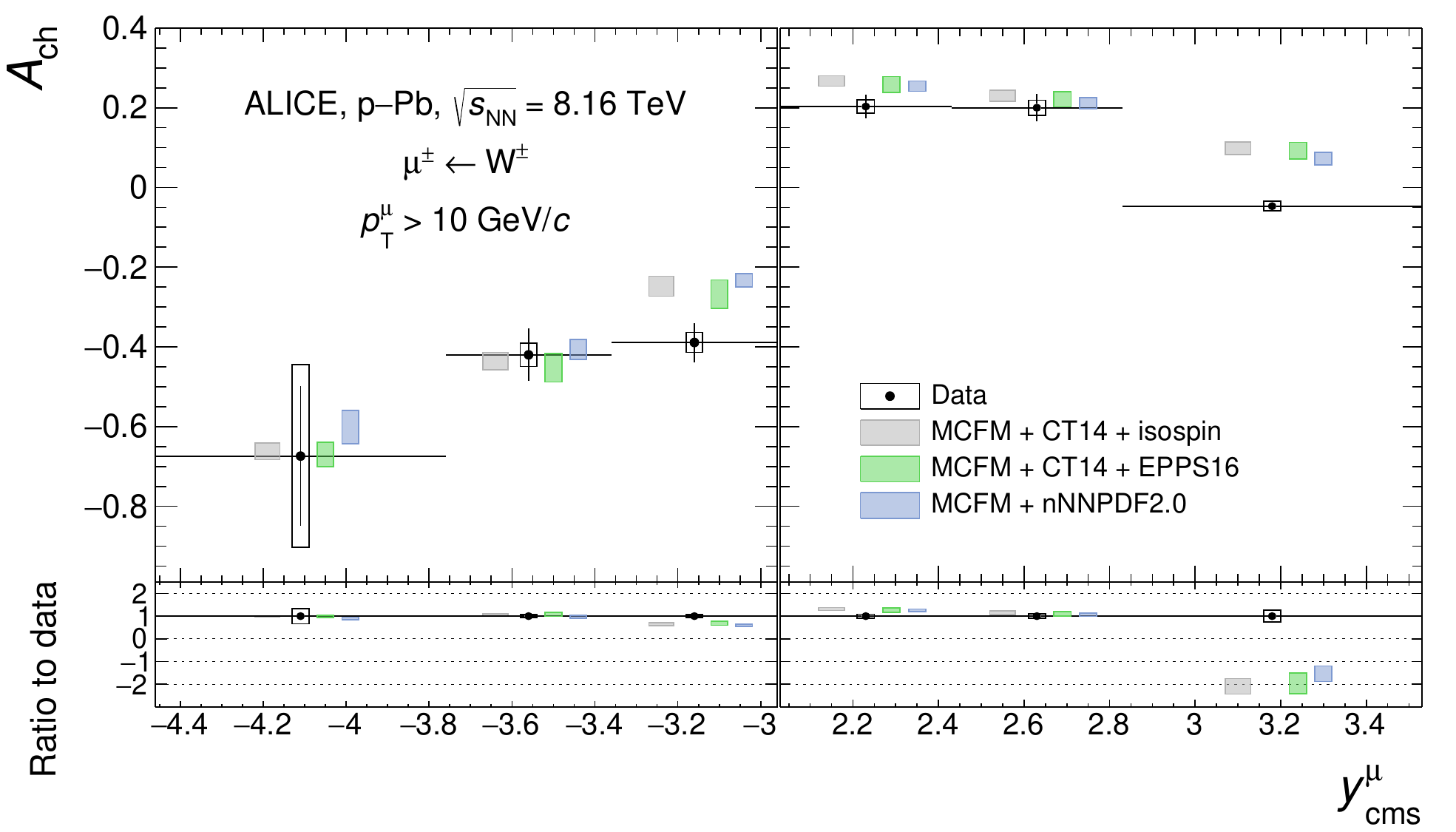}
    \end{center}
    \caption{Lepton charge asymmetry for muons from \W-boson decays with $\pt^\mu > 10$ \GeVc in \pPb collisions at \eightnn. The measurements are compared with predictions from pQCD calculations with several nPDF sets as well as with calculations based on the CT14 PDF~\cite{ct14} without nuclear modifications. All the calculations include the isospin effect. The bottom panels show the ratio of the calculated to the measured asymmetry. The horizontal bars correspond to the width of the rapidity intervals. The vertical bars and boxes indicate the statistical and systematic uncertainties, respectively. The data points are placed at the centres of the rapidity intervals while the theory points are horizontally shifted for better visibility.}
    \label{fig:pPb_Ach}
\end{figure}

The measured lepton charge asymmetries integrated over $\pt^\mu >$ 10 \GeVc in the rapidity intervals covered by the muon spectrometer for the two colliding beam configurations are:
\begin{equation*}
  A^{\text{Pb-going}}_{\rm ch} = -0.479 \pm 0.046 \text{ (stat)} \pm 0.056 \text{ (syst)}, \qquad
  A^{\text{p-going}}_{\rm ch} = 0.145 \pm 0.014 \text{ (stat)} \pm 0.021 \text{ (syst)}.
\end{equation*}
The measured $A_{\rm ch}$ as a function of rapidity is shown in Fig.~\ref{fig:pPb_Ach}. Consistently with the up and down quark compositions of the proton and Pb ion, the $A_{\rm ch}$ shows a predominance of \Wminus bosons at backward rapidities, in the Pb-going direction, and of \Wplus at forward rapidities. At very large positive rapidities, the lepton charge asymmetry becomes negative, which indicates a suppression of the \Wplus production. This suppression could be a consequence of the helicity conservation affecting the muonic decay of the boson, or a sharper slope of the up quark PDF in the shadowing region towards low Bjorken-$x$. The $A_{\rm ch}$ is compared with predictions from pQCD calculations with the CT14+EPPS16, nCTEQ15WZ, and nNNPDF2.0 PDFs sets, as well as with the CT14 PDF set for free nucleons. The calculations are performed at NLO and the same treatment of the isospin as for the production cross section is applied. The models reproduce the data well at backward rapidity, although a small tension is seen for the most central rapidity interval in which the theory predicts an increase of the charge asymmetry, while the measurement is independent of centrality within uncertainties. At forward rapidities, the model predictions and the measurements are in qualitative agreement, both showing a reduction of the charge asymmetry towards higher rapidities. However, the decrease seen in the data is notably larger than that in calculations, and it is interesting to note the sign inversion of the measured charge asymmetry, while the calculations stay positive over the whole forward rapidity interval. In the largest rapidity interval, the models are all lying more than 5$\sigma$ above the measurement.

    \subsubsection{Nuclear modification factor}

In \pPb collisions, the nuclear modification factor \RpPb, integrated over centrality, is calculated as
\begin{equation}
  \RpPb = \frac{1}{A} \times \frac{\dd \sigma^{\rm pPb}_{\W \rightarrow \muonpm \num} / \dd y^\mu_{\rm cms}}{\dd \sigma^{\rm pp}_{\W \rightarrow \muonpm \num} / \dd y^\mu_{\rm cms}}.
  \label{eqn:RpPb}
\end{equation}
It evaluates the deviation between the measured production cross section in \pPb collisions and the one expected from a superposition of uncorrelated pp collisions. It should be noted that for electroweak bosons, the \RpPb is a peculiar quantity. It is affected by the isospin effect, and as a consequence, expectation values for \RpPb can deviate from unity even in the absence of nuclear effects, such as the nuclear modification of the PDFs. Since no measurement of the \W-boson production in pp collisions at $\s~=~8.16$~TeV is available, the \RpPb presented here relies on theoretical calculations for the pp production cross section $\sigma^{\rm pp}_{\W \rightarrow \muonpm \num}$. The simulations are performed with the procedure discussed in Section~\ref{sec:analysis}, using POWHEG~\cite{powheg} interfaced with PYTHIA 6~\cite{pythia} for the event generation and CT10~\cite{ct10} for the proton PDF. It should be mentioned that the LHCb Collaboration has shown that the available models, including CT10, are able to describe well the production of \W bosons in pp collisions at similar rapidities and energies~\cite{Lhcb:Wpp8tev}. The associated uncertainty was evaluated by varying the strong coupling constant $\alpha_s$ within its uncertainties and using CTEQ6.6~\cite{cteq66} as an alternative PDF set, summing the sources in quadrature. The values of the \RpPb obtained for the \Wminus- and \Wplus-boson production integrated over $\pt^\mu >$ 10 \GeVc and the rapidity intervals covered by the muon spectrometer for the two colliding beam configurations are reported in Table~\ref{table:R-pPb}.

\begin{table}
  \centering
  \caption{Nuclear modification factors of the production of \Wminus and \Wplus bosons measured in their muonic decays in \pPb collisions at \eightnn, for muons with $\pt^\mu > 10$ \GeVc. The pp reference cross sections are taken from simulations using the POWHEG~\cite{powheg} generator and CT10 PDF~\cite{ct10}. The quoted uncertainties correspond to the statistical and systematic uncertainties on the \pPb measurement, and to the asymmetric systematic uncertainty on the pp reference, respectively.}
  \renewcommand{\arraystretch}{1.5}
  \begin{tabular}{|c|c|c|}
    \cline{2-3}
    \multicolumn{1}{c|}{}               & Pb-going ($-4.46 < y^\mu_{\rm cms} < -2.96$)         & p-going ($2.03 < y^\mu_{\rm cms} < 3.53$) \\
    \hline
    \RpPb ($\Wminus \rightarrow \muon \Anum$) & $1.620 \pm 0.057 \pm 0.079 ^{+ 0.092}_{- 0.062}$ & $0.888 \pm 0.047 \pm 0.080 ^{+ 0.060}_{- 0.039}$ \\
    \RpPb ($\Wplus \rightarrow \Amuon \num$)  & $0.643 \pm 0.036 \pm 0.051 ^{+ 0.046}_{- 0.031}$ & $0.793 \pm 0.034 \pm 0.051 ^{+ 0.048}_{- 0.037}$ \\
    \hline
  \end{tabular}
  \renewcommand{\arraystretch}{1.5}
  \label{table:R-pPb}
\end{table}

\begin{figure}[!h]
    \begin{center}
    \includegraphics[width = 0.90\textwidth]{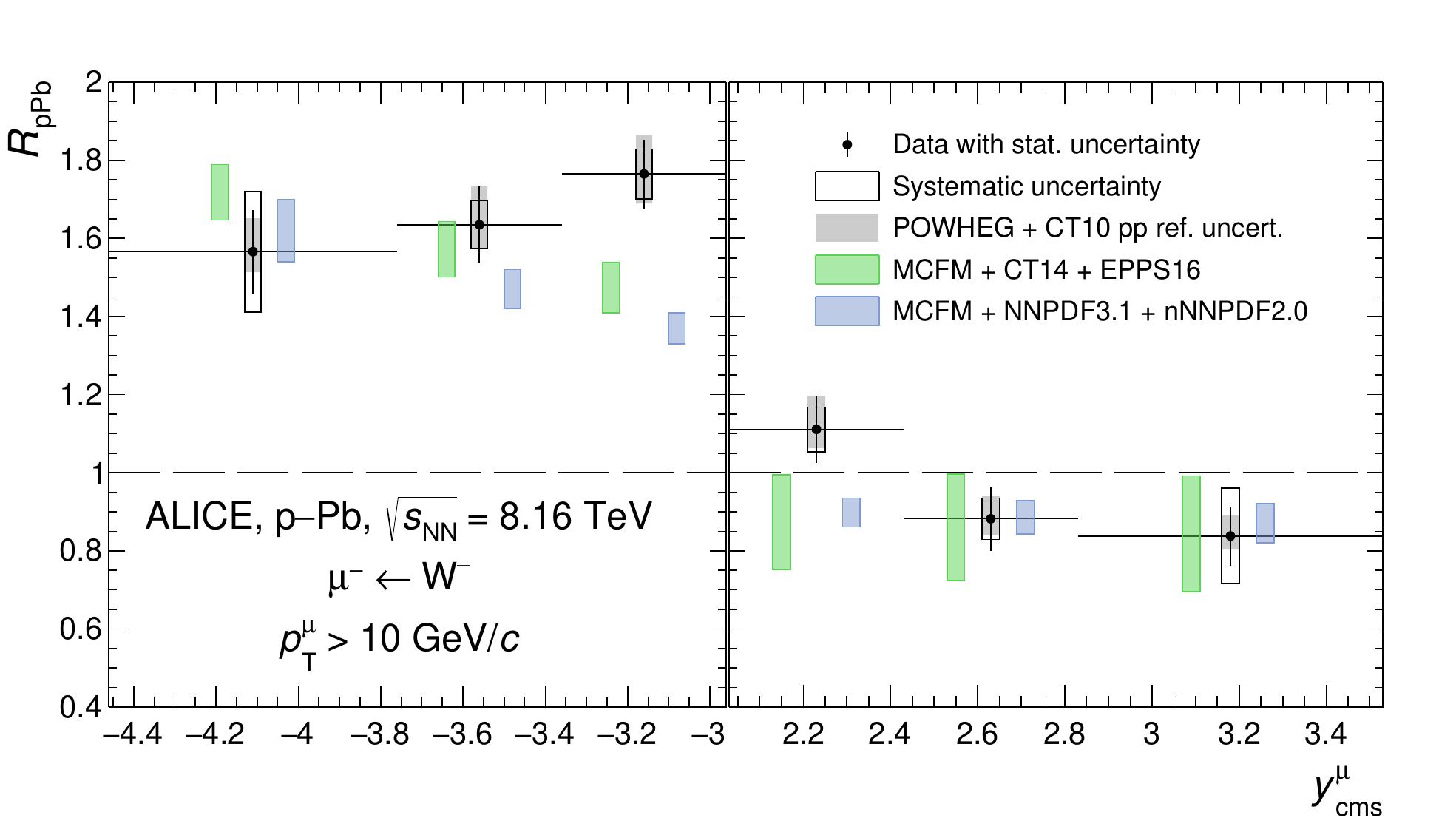}
    \includegraphics[width = 0.90\textwidth]{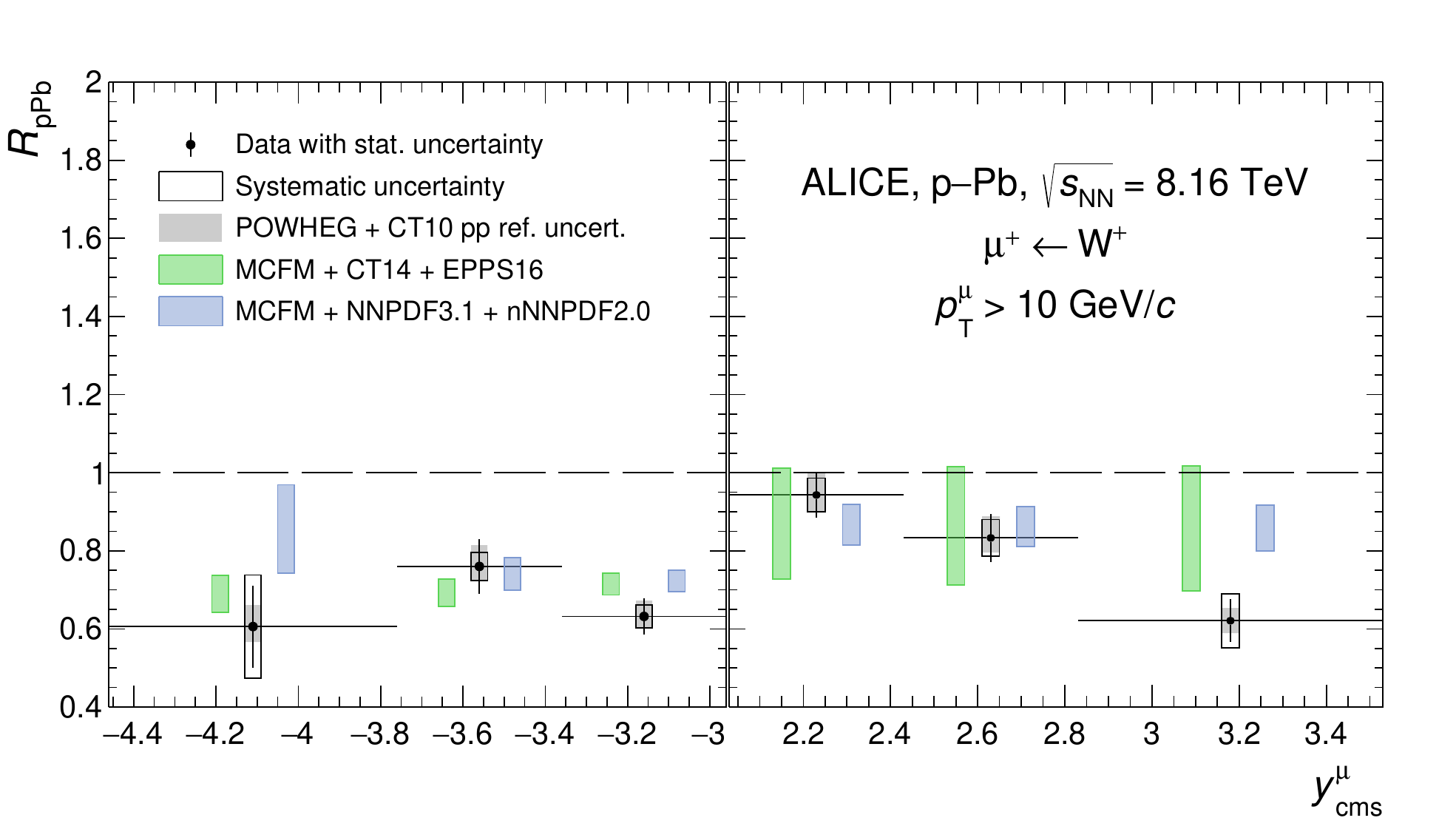}
    \end{center}
    \caption{Nuclear modification factor \RpPb for muons from \Wminus (top) and \Wplus (bottom) decays with $\pt^\mu~>$~10~\GeVc in \pPb collisions at \eightnn. The measurements are compared with predictions from pQCD calculations with several nPDF sets. The horizontal bars correspond to the width of the rapidity bins. The vertical bars and boxes indicate the statistical and systematic uncertainties respectively. The grey bands indicate the uncertainty on the pp production cross section. The data points are placed at the centres of the rapidity intervals while the theory points are horizontally shifted for better visibility.}
    \label{fig:pPb_RpA}
\end{figure}

The measured \RpPb is shown in Fig.~\ref{fig:pPb_RpA} as a function of rapidity,
where it is compared with predictions from the same models, and obtained using the same framework, as for the asymmetry $A_{\rm ch}$. It should be noted that the nNNPDF2.0 predictions rely on a different baseline PDF, employing NNPDF3.1~\cite{nnpdf} instead of the CT14 model used in the calculations with EPPS16 nPDFs. For both charges of the boson, the measured \RpPb is independent of $y$ at backward rapidities, within the uncertainties. This trend is satisfactorily reproduced by the models for the \Wplus boson. For the \Wminus boson however, a significant rapidity dependence is seen in the two calculations, which underestimate the measured value in the most central interval. At forward rapidities, the measured \RpPb decreases for the two charges while the calculations show a flat behaviour. This creates a tension with the calculations in the intervals closest to midrapidity for the \Wminus boson, where the models underestimate the measurement, and in the largest rapidity interval for the \Wplus boson where the nNNPDF2.0 model overestimates the data.

    \subsubsection{Production as a function of the collision centrality}

The production of muons from \W-boson decays is studied as a function of the collision centrality. Electroweak-boson production occurs in hard scattering processes, during the initial stages of the collision, and is expected to scale with the number of binary nucleon--nucleon collisions, provided that the evaluation of the centrality is unbiased. As mentioned in Section~\ref{sec:data}, and in order to avoid the bias in multiplicity-based centrality estimators, the classification in centrality intervals is performed based on the energy deposited by the spectator (non-interacting) nucleons in the neutron zero-degree calorimeters (ZN) in the Pb-going side. The study of the centrality dependence of the \W-boson yield can therefore also serve as a test bench for the centrality estimation.

\begin{figure}
    \begin{center}
    \includegraphics[width = 0.49\textwidth]{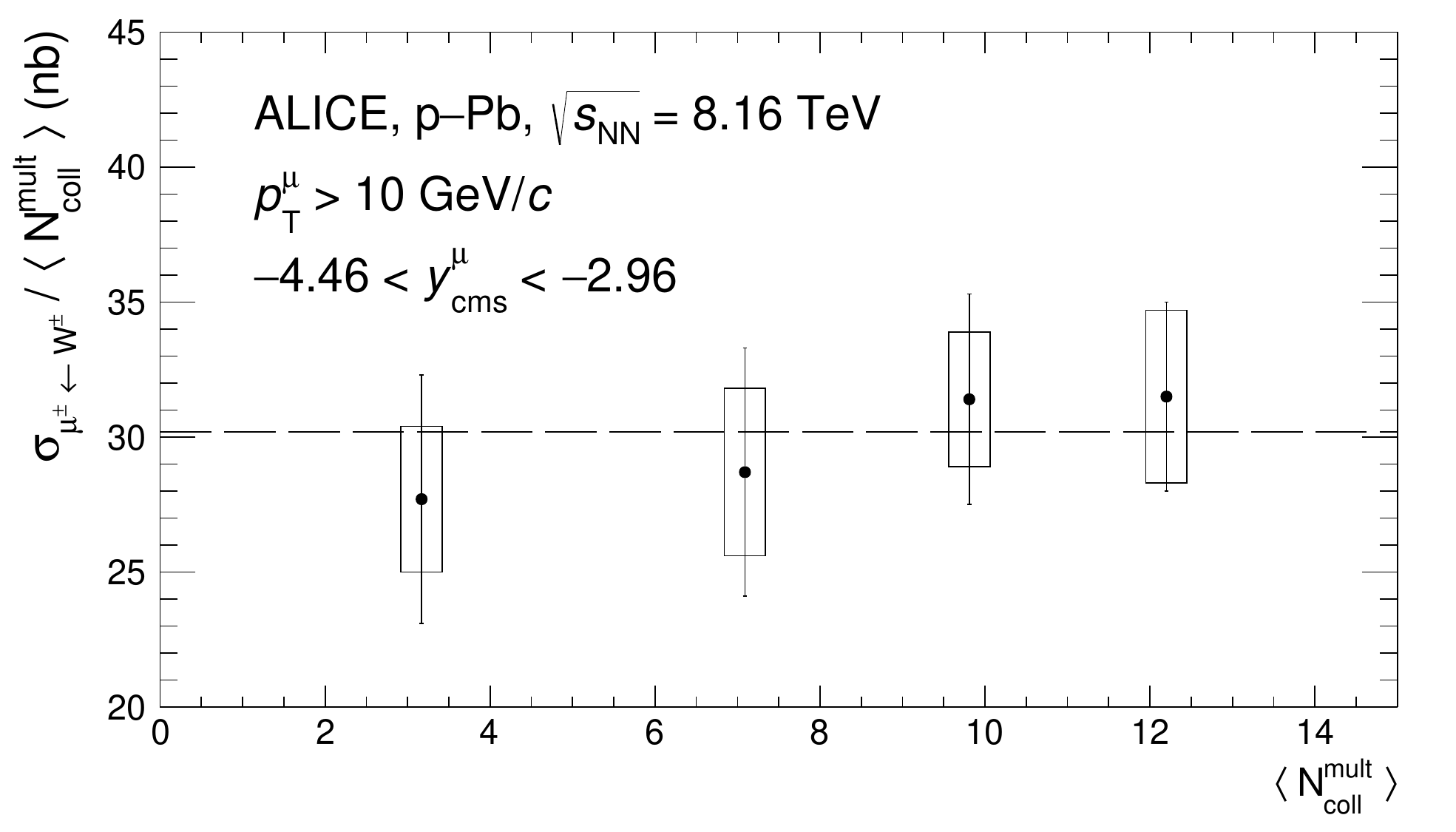}
    \includegraphics[width = 0.49\textwidth]{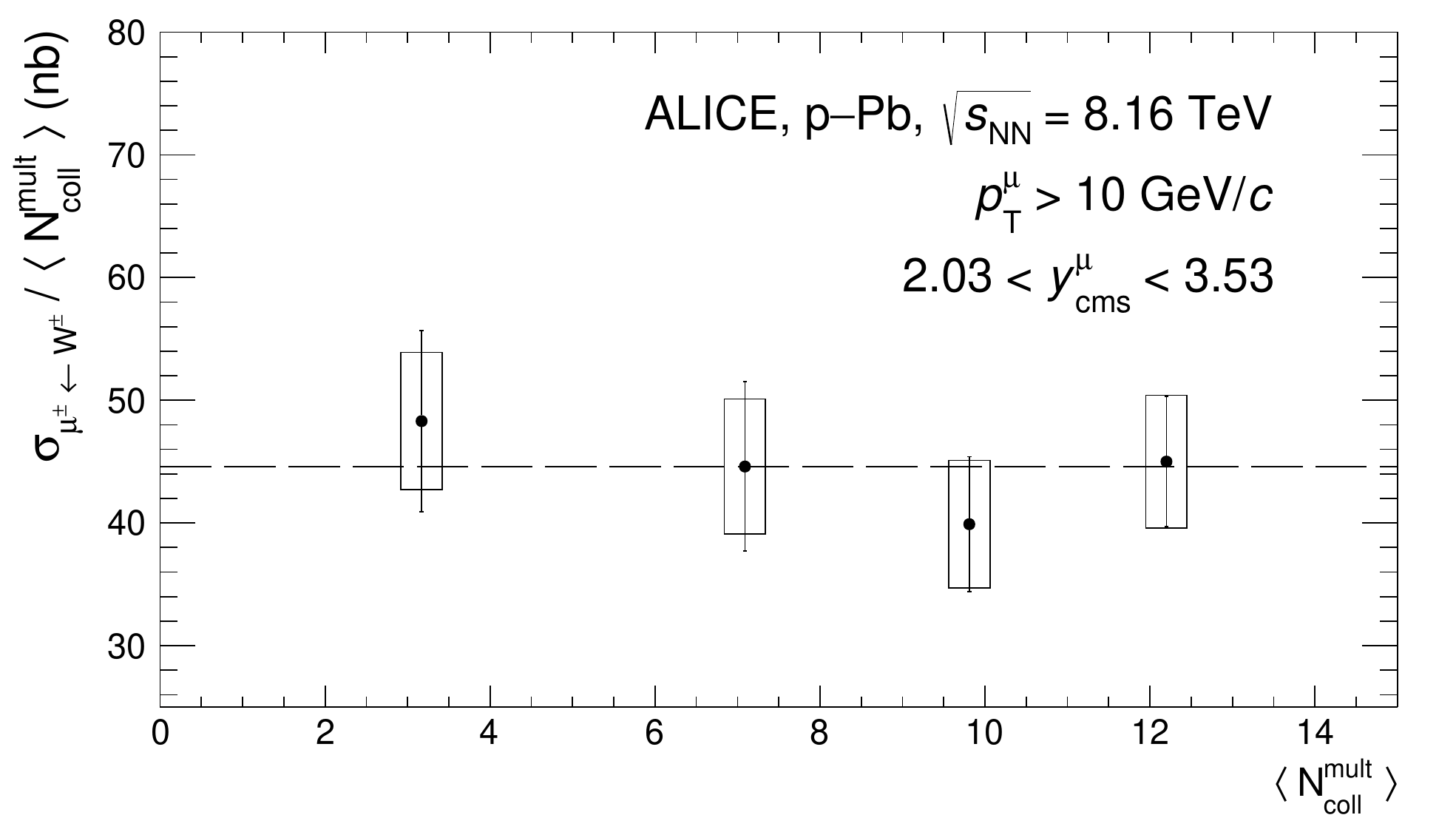}
    \end{center}
    \caption{Combined yield of muons from \Wminus and \Wplus decays with $\pt^\mu >$ 10 \GeVc, normalised by the average number of binary nucleon--nucleon collisions \avNcollMult in \pPb collisions at \eightnn in the Pb-going (left) and p-going (right) configurations. The bars and boxes correspond to statistical and systematic uncertainties respectively. The horizontal dashed line indicates the central value of the yield normalised to \avNcollMult measured for the 0--100\% centrality interval.}
    \label{fig:pPb_centrality}
\end{figure}

In order to maximise the amount of signal in each centrality class, the \Wminus and \Wplus yields are combined. The cross section normalised to the average number of nucleon--nucleon collisions, \avNcollMult, is then calculated as
\begin{equation}
  \frac{1}{\avNcollMult} \times \frac{N^i_{\W}}{\lumi_{\rm int} \times f^i_{\rm MB} \times \epsilon},
\end{equation}
where \avNcollMult is the average number of binary nucleon--nucleon collisions, $N^i_{\W}$ is the number of muons from \W decays in a given centrality class $i$, and $f^i_{\rm MB}$ is the fraction of MB-triggered events in the centrality class $i$ to those in the full centrality range (0--100\%). The cross sections for the two colliding beam configurations, normalised to \avNcollMult and averaged over centrality, amount to:
\begin{equation*}
    \begin{split}
    -4.46 < y^\mu_{\rm cms} < -2.96: & \qquad \sigma_{\muonpm \leftarrow \W} / \avNcollMult = 30.2 \pm 2.0 \text{ (stat)} \pm 2.8 \text{ (syst)} \text{ nb}, \\
    2.03 < y^\mu_{\rm cms} < 3.53:   & \qquad \sigma_{\muonpm \leftarrow \W} / \avNcollMult = 44.6 \pm 3.3 \text{ (stat)} \pm 5.1 \text{ (syst)} \text{ nb}. \\
    \end{split}
\end{equation*}
The normalised cross sections are shown as a function of \avNcollMult in Fig.~\ref{fig:pPb_centrality}. The horizontal dashed line in the figure indicates the central value of the centrality-averaged measurement. The measured yield divided by \avNcollMult is found to be independent of centrality within uncertainties.

  \subsection{\PbPb collisions}

    \subsubsection{Production cross sections and lepton charge asymmetry}

\begin{figure}[h]
    \begin{center}
    \includegraphics[width = 0.90\textwidth]{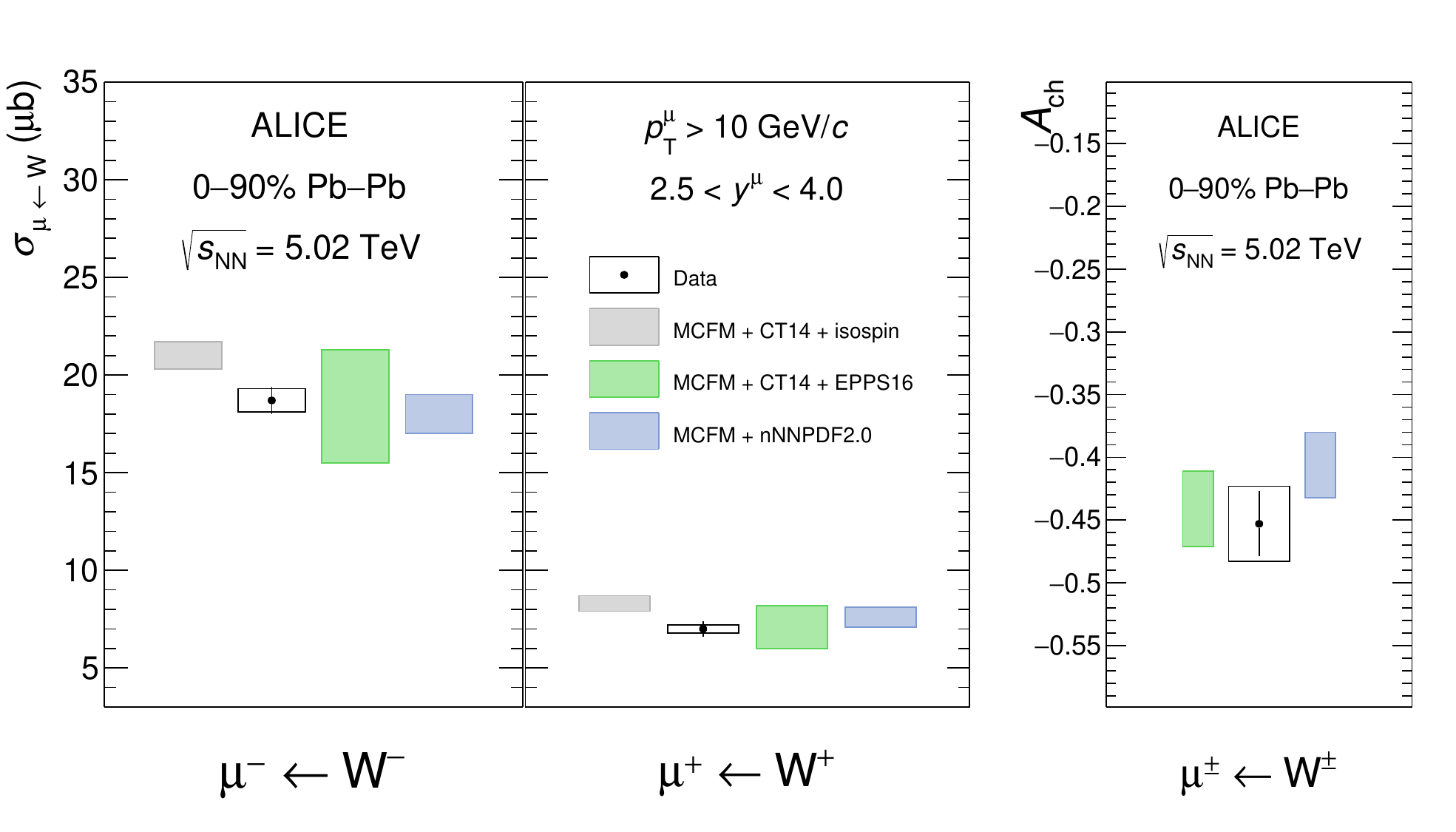}
    \end{center}
    \caption{Production cross section (left panel) and lepton charge asymmetry (right panel) of the $\W \rightarrow \muonpm \num$ processes for the 0--90\% centrality class, for muons with $\pt^\mu > 10$ \GeVc and $2.5 < y^\mu_{\rm cms} < 4.0$, in \PbPb collisions at \fivenn. The measured cross sections and the asymmetry are compared with predictions using the CT14+EPPS16~\cite{ct14,epps16} combination, nNNPDF2.0~\cite{nnnpdf} nPDF model, as well as calculations with the CT14~\cite{ct14} PDF without nuclear corrections. All the calculations include the isospin effect. The vertical bars and boxes around the data points indicate the statistical and systematic uncertainties, respectively.}
    \label{fig:PbPb_xSecAch}
\end{figure}

In \PbPb collisions at \fivenn, the production cross section and lepton charge asymmetry of muons from \W-boson decays are evaluated as in \pPb collisions, from Eqs.~\ref{eqn:xSec} and~\ref{eqn:Ach}. The V0M amplitude is used to estimate the centrality of the collision. The production cross sections for \Wminus and \Wplus bosons in the 0--90\% centrality class are
\begin{equation*}
  \sigma_{\Wminus \rightarrow \muon \Anum} = 18.7 \pm 0.7 \text{ (stat)} \pm 0.6 \text{ (syst)} \; \mu\text{b}, \qquad
  \sigma_{\Wplus \rightarrow \Amuon \num} = 7.0 \pm 0.4 \text{ (stat)} \pm 0.2 \text{ (syst)} \; \mu\text{b}.
\end{equation*}
In the left panel of Fig.~\ref{fig:PbPb_xSecAch}, these values are compared with pQCD calculations using the CT14~\cite{ct14}, the CT14+EPPS16 combination~\cite{ct14,epps16}, and nNNPDF2.0~\cite{nnnpdf} PDF sets, all accounting for the isospin of the \PbPb system. In the \PbPb collision system, one cannot disentangle the high and low Bjorken-$x$ ranges, as it was possible in \pPb collisions. The comparison of the production for positively and negatively charged bosons shows the effect of the isospin, since the up- and down-quark densities in the Pb nucleus favour the production of \Wminus and suppress that of \Wplus. The measured cross sections are lower than the predictions with the CT14 PDFs for free nucleons, suggesting a significant effect due to nuclear modifications of the PDFs on the \W-boson production in \PbPb collisions. The calculations including the EPPS16 and nNNPDF2.0 nuclear modifications are consistent with the data within uncertainties. As it was observed in the predictions in \pPb collisions, the nNNPDF2.0 calculations have significantly smaller uncertainties.

The lepton charge asymmetry in the 0--90\% centrality interval is measured to be
\begin{equation*}
  A_{\rm ch} = -0.453 \pm 0.026 \text{ (stat)} \pm 0.030 \text{ (syst)}.
\end{equation*}
In the right panel of Fig.~\ref{fig:PbPb_xSecAch}, this observable is compared with pQCD calculations using the CT14+ EPPS16~\cite{epps16} and nNNPDF2.0~\cite{nnnpdf} nPDFs. Both models describe well the measured value. The partial cancellation of uncertainty in the $A_{\rm ch}$ has a remarkable effect on the EPPS16 prediction, as the theoretical uncertainties are now similar to that in the data and nNNPDF2.0 calculation. \\

    \subsubsection{Normalised yield as a function of the collision centrality}

\begin{figure}
    \begin{center}
    \includegraphics[width = 0.49\textwidth]{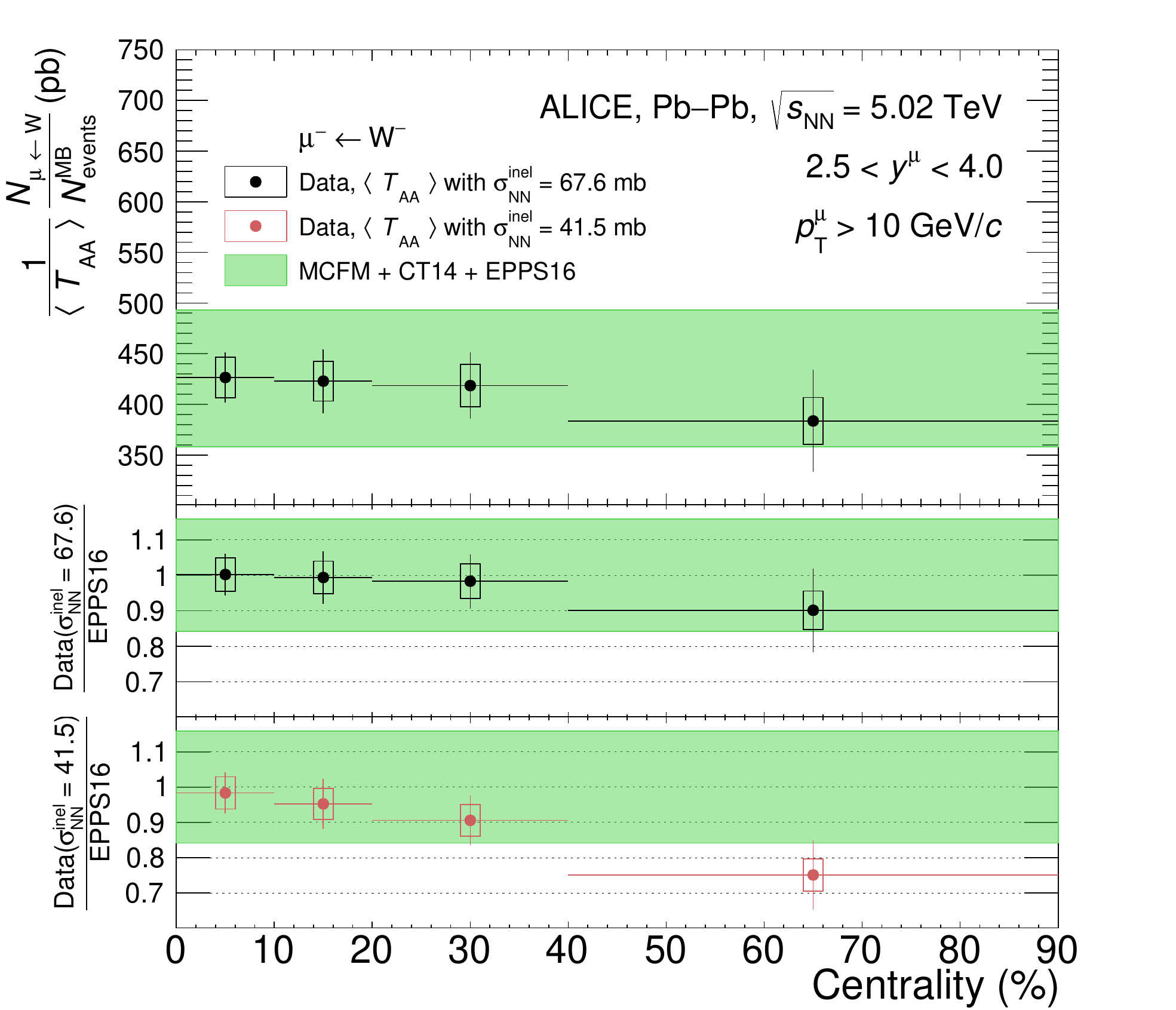}
    \includegraphics[width = 0.49\textwidth]{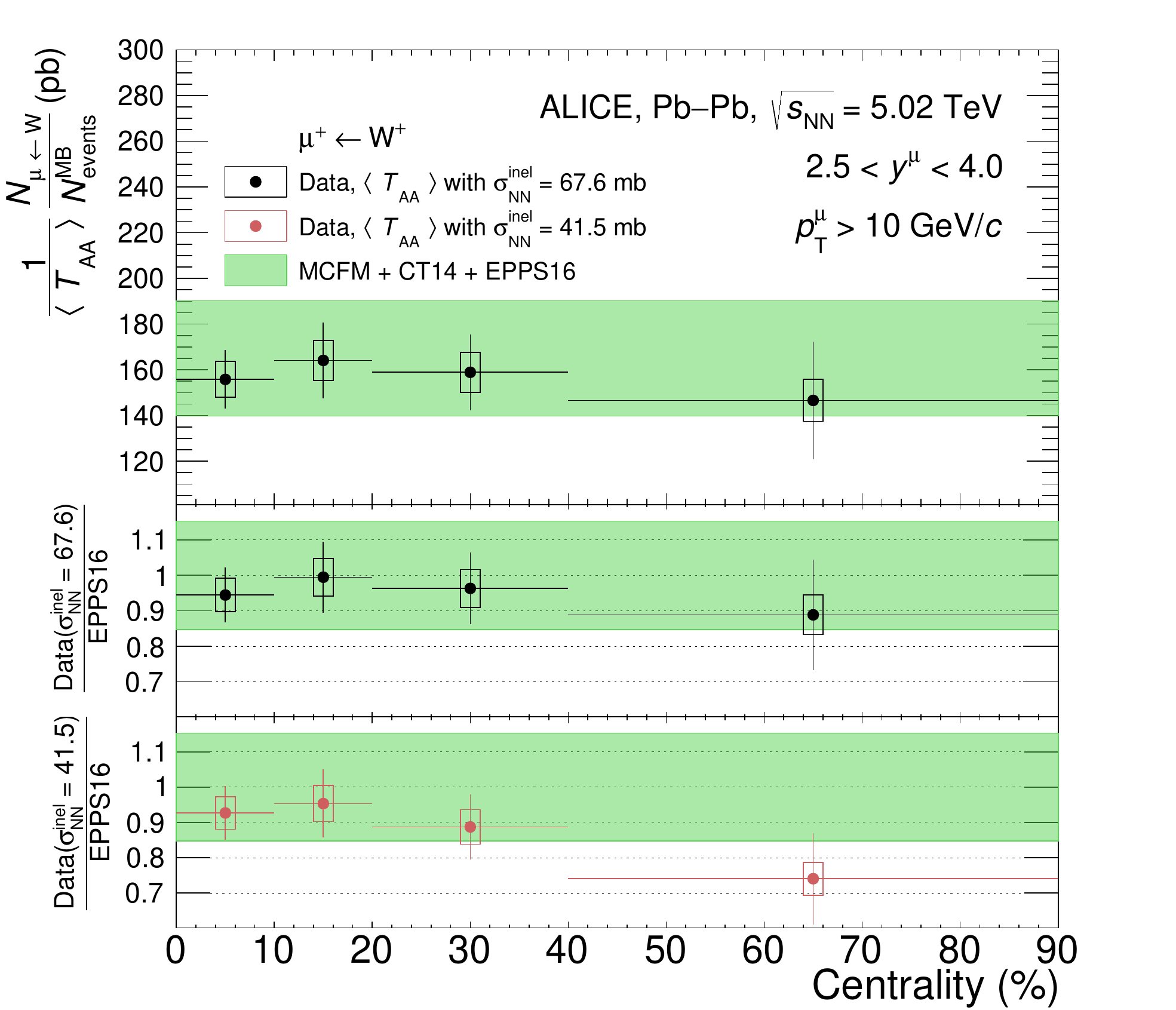}
    \end{center}
    \caption{\avTaa-scaled yield of muons from \Wminus (left) and \Wplus (right) decays in \PbPb collisions at \fivenn for muons with $\pt^\mu > 10$ \GeVc and $2.5 < y^\mu_{\rm cms} < 4.0$. In the top panels the yield is compared with pQCD calculations using the CT14 PDF~\cite{ct14} as baseline and implementing the EPPS16~\cite{epps16} nuclear modifications. The ratio to theory of the measured yield normalised with \avTaa evaluated with $\sigma^{\rm inel}_{\rm NN}$ = 67.6 mb and 47.5 mb is shown in the middle and bottom panels, respectively (see the text for details). The horizontal bars indicate the width of the centrality intervals, the vertical bars and boxes correspond to the statistical and systematic uncertainties, respectively. The band indicate the uncertainty on the theoretical computations.}
    \label{fig:PbPb_yield}
\end{figure}

The normalised yield is obtained by dividing the yield of muons from \W decays, $N_{\muonpm \leftarrow \W}$, by the equivalent number of MB events $N^{\rm MB}_{\rm events}$, and then normalising to the average nuclear overlap function \avTaa~\cite{centrality}:
\begin{equation}
  \frac{1}{\avTaa} \times \frac{N_{\muonpm \leftarrow \W}}{N^{\rm MB}_{\rm events}}.
\end{equation}
In the 0--90\% centrality class, the binary-scaled yield amounts to
\begin{equation*}
    \begin{split}
  N_{\muon \leftarrow \Wminus} /\left( N^{\rm MB}_{\rm events} \times \avTaa \right) &= 420.5 \pm 16.4 \text{ (stat)} \pm 18.0 \text{ (syst)} \text{ pb}, \\
  N_{\Amuon \leftarrow \Wplus} /\left( N^{\rm MB}_{\rm events} \times \avTaa \right) &= 158.5 \pm 8.2 \text{ (stat)} \pm 6.9 \text{ (syst)} \text{ pb}.
    \end{split}
  \end{equation*}
The \W-boson yield normalised to \avTaa as a function of the collision centrality is shown in Fig.~\ref{fig:PbPb_yield} for both charges of the boson. The \avTaa-scaled yields are independent of centrality, as expected from the binary scaling of \W-boson production in nuclear collisions assuming negligible centrality dependence of the shadowing. The measurents are compared with pQCD calculations using the CT14~\cite{ct14} PDF combined with the EPPS16~\cite{epps16} nuclear modifications. A good agreement with the theory is found for both charges of the boson.

The centrality dependence of the PDF modifications has been explored through impact-parameter dependent nPDFs~\cite{emelyanov1999,helenius2012}, but calculations of electroweak-boson production within this approach show a very limited dependence on the centrality, as reported in Ref.~\cite{Alice:ZpPb8tevPbPb5tev}. A possible centrality dependence of the production in terms of shadowing of the inelastic nucleon--nucleon cross section $\sigma^{\rm inel}_{\rm NN}$ was proposed in Ref.~\cite{epps-new}. In that study, the standard paradigm of extracting $\sigma^{\rm inel}_{\rm NN}$ from pp data, is questioned as a potential source of bias. The re-evaluation of the inelastic cross section from ATLAS measurements of electroweak-boson production in \PbPb collisions~\cite{Atlas:WPbPb5tev,Atlas:ZPbPb5tev} yields $\sigma^{\rm inel}_{\rm NN} = 41.5^{+16.2}_{-12.0} \text{ mb}$, a value significantly lower than the one used for centrality determination in \PbPb collisions at \fivenn with ALICE, taken as $\sigma^{\rm inel}_{\rm NN} = 67.6 \pm 0.6$ mb~\cite{centrality}. This alternative value of the inelastic cross section is found to improve the agreement between the ATLAS data and the pQCD calculations.

The bottom panels of Fig.~\ref{fig:PbPb_yield} show the centrality-dependent measurements obtained by normalising the yield with \avTaa evaluated using the nuclear-suppressed inelastic cross section from Ref.~\cite{epps-new}. The distributions show a significant centrality dependence, with the \Wminus distribution deviating from the binary scaling. This alternative value of the inelastic cross section, which provides a better agreement between pQCD calculations and the ATLAS measurement in peripheral collisions, has the opposite effect here. The yield normalised to \avTaa with $\sigma^{\rm inel}_{\rm NN} = 41.5$ mb shows a worse agreement with the theory than that with $\sigma^{\rm inel}_{\rm NN} = 67.5$ mb in the 40--90\% centrality interval. The authors of Ref.~\cite{epps-new} expect other effects to be possibly relevant in peripheral collisions, such as a possible centrality dependence of $\sigma^{\rm inel}_{\rm NN}$ and the neutron-skin effect, which could explain the tension with the data for peripheral collisions. It should be noted that the neutron skin effect would affect the production of \Wminus and \Wplus bosons in opposite directions, enhancing the former and suppressing the latter, thus not substantially improving the description of the measurements. \\

\begin{figure}
    \begin{center}
    \includegraphics[width=0.9\textwidth]{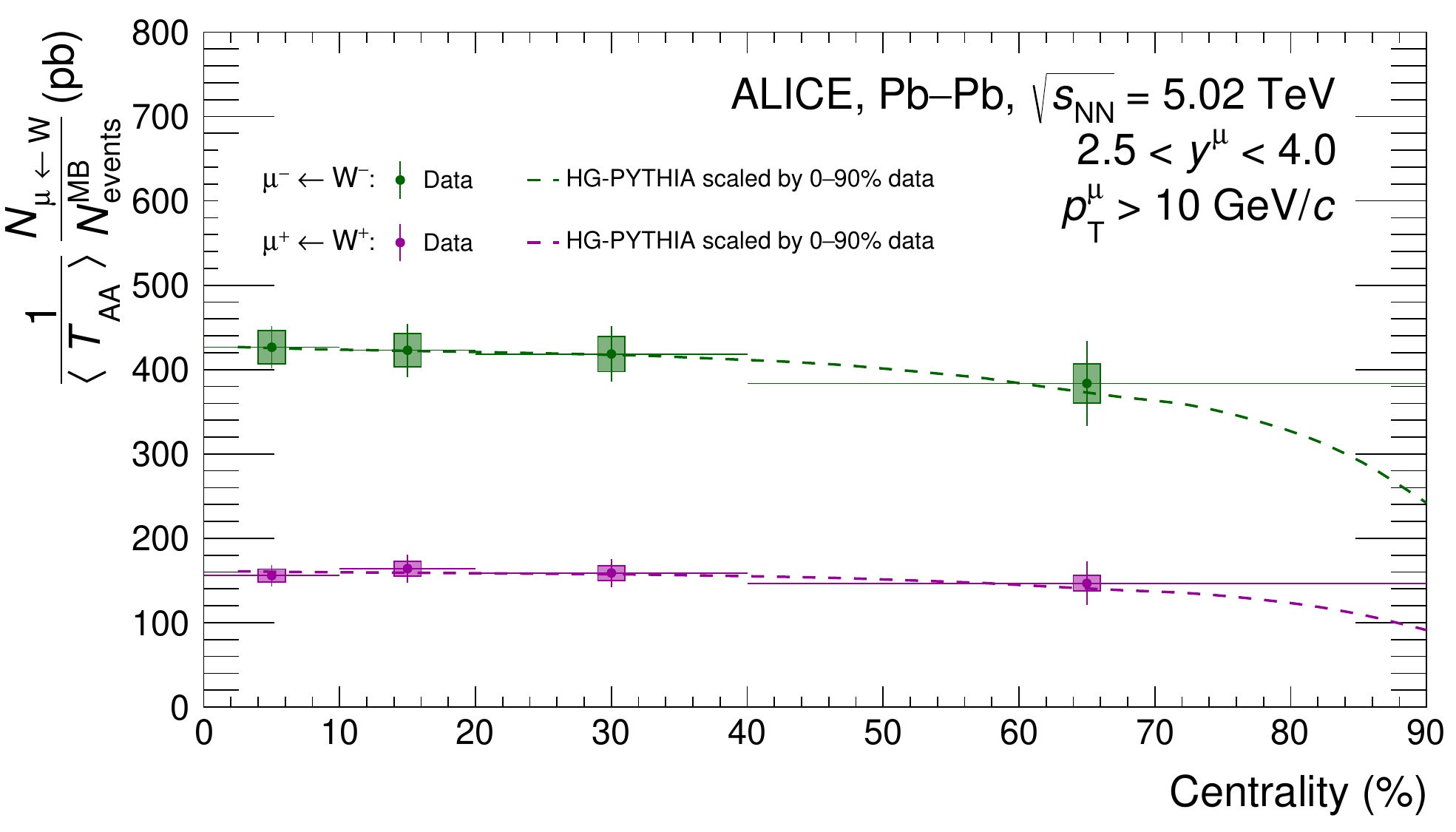}
    \end{center}
    \caption{\avTaa-scaled yield of muons from \Wminus and \Wplus decays in \PbPb collisions at \fivenn for muons with $\pt^\mu > 10$ \GeVc and $2.5 < y^\mu_{\rm cms} < 4.0$. The measured production is compared with HG-PYTHIA~\cite{hg-pythia} calculations of the \RPbPb of hard scatterings scaled with the centrality-averaged production in 0--90\% centrality, indicated as dashed lines. The horizontal bars correspond to the width of the centrality intervals, the vertical bars and boxes indicate the statistical and systematic uncertainties, respectively.}
    \label{fig:PbPb_hg-pythia}
\end{figure}

Recent measurements of the \avTaa-normalised yield of the \Z boson~\cite{Atlas:ZPbPb2tev,Atlas:ZPbPb5tev,Cms:ZPbPb2tev-2,Cms:ZPbPb5tev} have shown a decreasing trend for the most peripheral events, contradicting the binary-scaling assumption. This phenomenon has also been observed and studied by the ALICE Collaboration~\cite{alice-peripheral} for charged particle production. A possible explanation for this observation has been formulated in terms of event selection and geometry biases affecting peripheral events in the HG-PYTHIA model~\cite{hg-pythia}. In order to compare it with the \PbPb measurements presented in this article, the \RPbPb for hard scatterings calculated with this model was scaled by the centrality-averaged, \avTaa-normalised yields of \Wminus and \Wplus bosons measured in the 0--90\% centrality class. The resulting distributions are compared with the centrality-dependent measurements in Fig.~\ref{fig:PbPb_hg-pythia}. The scaled calculations are in good agreement with the data, although the small \W yield in peripheral collisions does not allow for a granularity fine enough in the 40--90\% centrality interval to show, if any, a statistically significant decrease of the production in this region.

\begin{figure}
    \begin{center}
    \includegraphics[width = 0.90\textwidth]{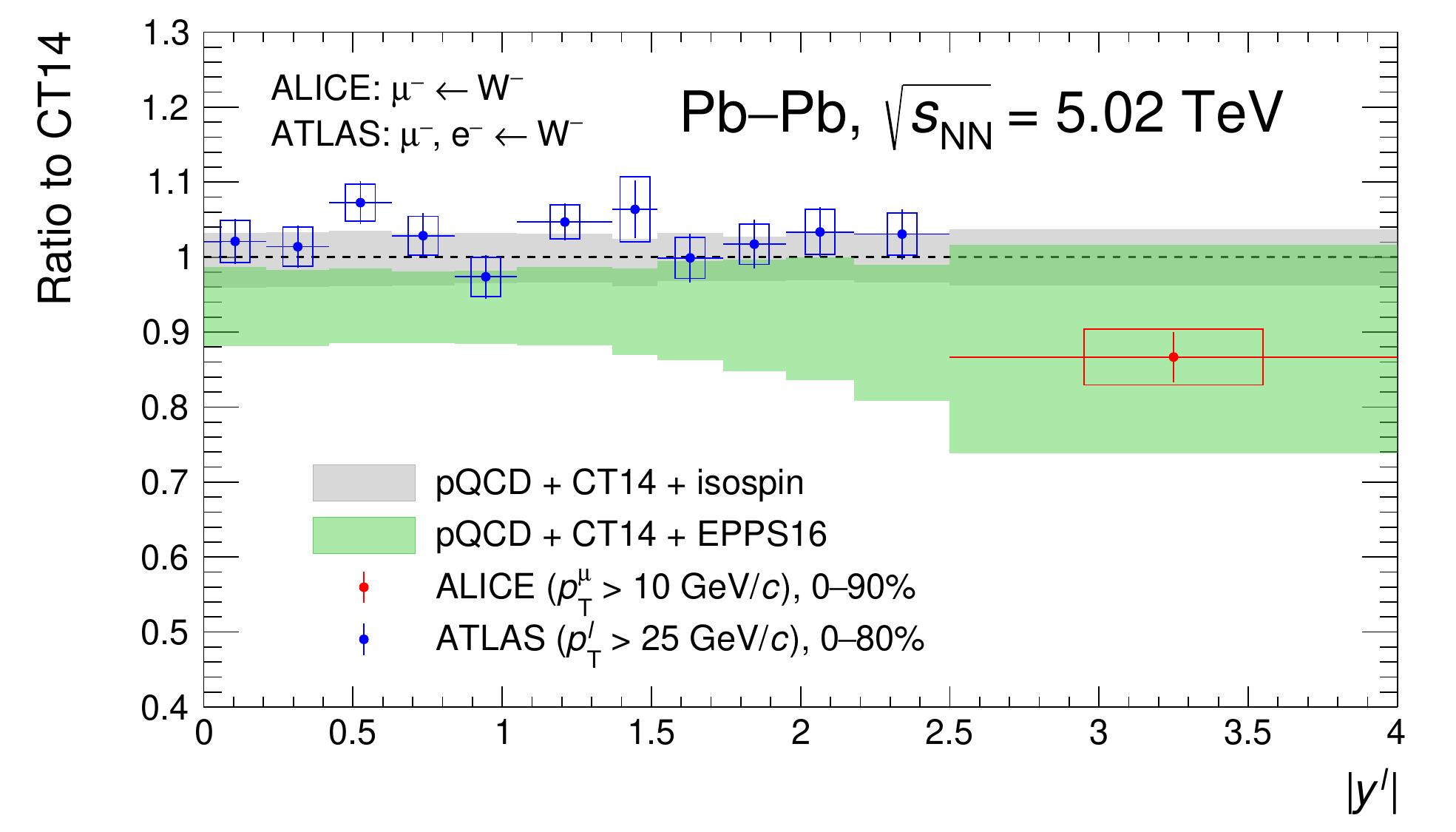}
    \includegraphics[width = 0.90\textwidth]{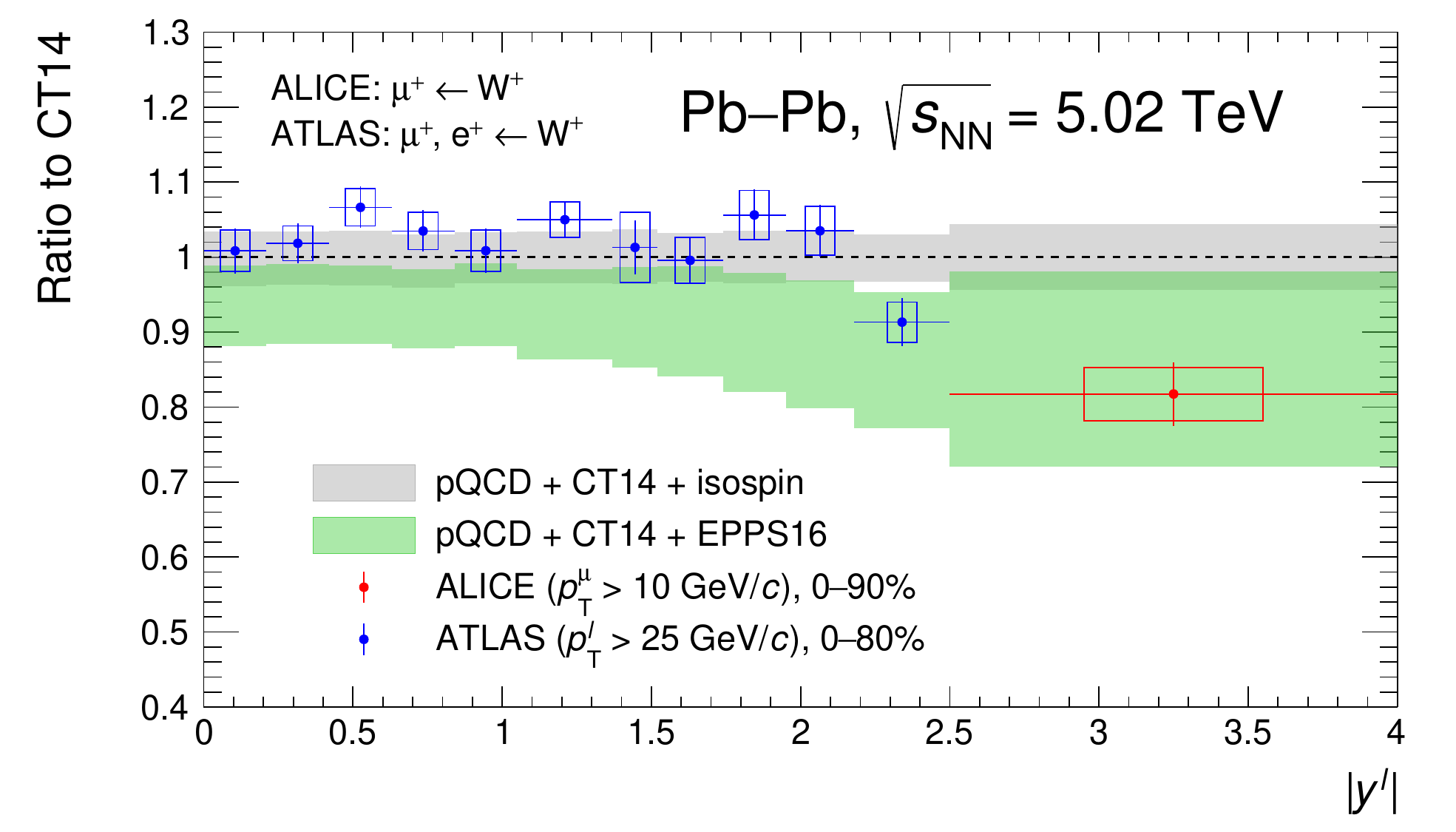}
    \end{center}
    \caption{Ratio to pQCD calculations with CT14 PDFs~\cite{ct14} of the production of muons from \Wminus (top) and \Wplus (bottom) decays measured as a function of rapidity in \PbPb collisions at \fivenn by the ALICE and ATLAS~\cite{Atlas:WPbPb5tev} Collaborations. The ratio of EPPS16+CT14~\cite{epps16} calculations to that of CT14-only calculations is also shown. The grey band around the line at unity indicates the uncertainty on the calculations with CT14 PDFs.}
    \label{fig:PbPb_compATLAS}
\end{figure}

The ATLAS Collaboration measured the production of \W bosons in the electronic and muonic decay channels in \PbPb collisions at \fivenn~\cite{Atlas:WPbPb5tev}. Their results are reported for the 0--80\% centrality class and are extracted from a data sample corresponding to a total integrated luminosity of 0.49 nb$^{-1}$. The decay leptons are detected in the rapidity interval $|y| < 2.5$, allowing for a complete continuity with the ALICE measurement in $2.5 < y < 4.0$. Similarly to the CMS measurements presented in Section~\ref{sec:pPb}, the ATLAS Collaboration also applied a tighter selection on the lepton \pt, at 25 \GeVc, the comparison is thus performed by means of the ratio between the measured \W-boson yields and the predictions from two pQCD calculations, the first using the EPPS16~\cite{epps16} nPDF set and the second using the CT14~\cite{ct14} PDFs. The comparison as a function of rapidity is shown in the two panels of Fig.~\ref{fig:PbPb_compATLAS} for the two charges of the boson.

The ALICE measurements are lower by 2$\sigma$ than the CT14 predictions and are described by EPPS16. The ATLAS data, instead, are better described by calculations without nPDF effects. This comparison motivated the study in Ref.~\cite{epps-new} with a shadowing-induced reduction of the inelastic cross section, but other possible origins of the effect have also been proposed~\cite{jonas2021}.

    \subsubsection{Nuclear modification factor}

In the \PbPb analysis, the nuclear modification factor of muons from \W-boson decays is evaluated by dividing the \avTaa-scaled yield by the \W-boson production cross section in pp collisions:
\begin{equation}
  R_{\rm AA} = \frac{1}{\avTaa} \times \frac{N^{\rm MB}_{\muonpm \leftarrow \W}}{\sigma^{\muonpm \leftarrow \W}_{\rm pp}},
\end{equation}
where $N^{\rm MB}_{\muonpm \leftarrow \W}$ is the number of muons from \W decays per MB event, $\sigma^{\muonpm \leftarrow \W}_{\rm pp}$ is the $\muonpm \leftarrow \W$ cross section in pp collisions, and $\avTaa$ is the average nuclear overlap function for the considered centrality class. As in \pPb collisions, the pp production cross section and the associated uncertainty were obtained from POWHEG and PYTHIA 6~\cite{powheg, pythia} simulations using CT10~\cite{ct10} for the proton PDF. For the 0--90\% centrality interval, the $R_{\rm AA}$ of muons from \Wminus- and \Wplus-boson decays are:
\begin{equation*}
\begin{split}
  R^{\muon \leftarrow \Wminus}_{\rm AA} = 1.32 \pm 0.05 \text{ (stat)} \pm 0.06 \text{ (syst)} \pm 0.14 \text{ (pp ref.)}, \\
  R^{\Amuon \leftarrow \Wplus}_{\rm AA} = 0.57 \pm 0.03 \text{ (stat)} \pm 0.02 \text{ (syst)} \pm 0.07 \text{ (pp ref.)}.
\end{split}
\end{equation*}
The production of \Wminus is enhanced, and that of \Wplus is suppressed relative to pp collisions, as expected following the content in u and d quarks of the Pb nucleus.

\begin{figure}
    \begin{center}
    \includegraphics[width = 0.90\textwidth]{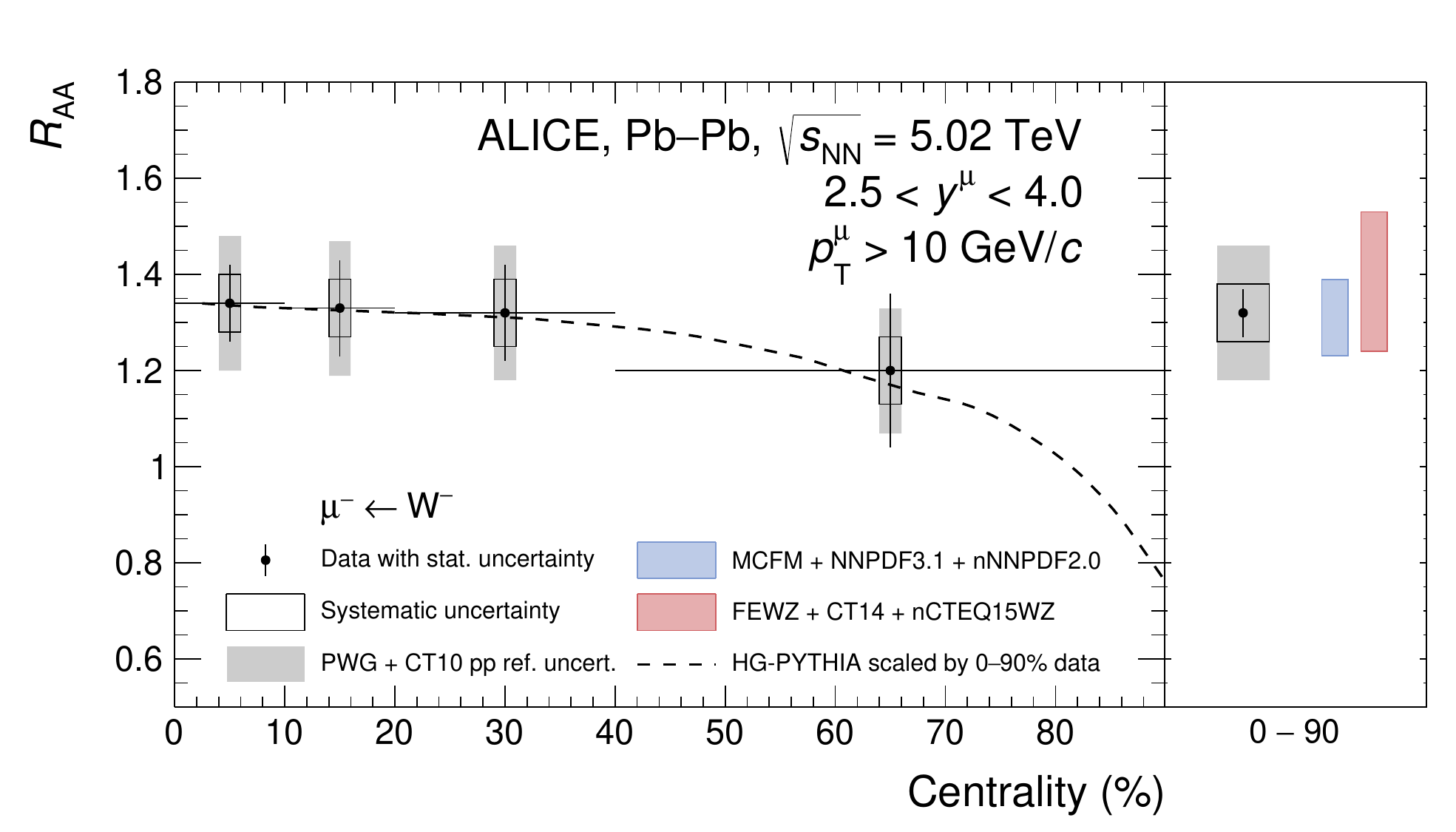}
    \includegraphics[width = 0.90\textwidth]{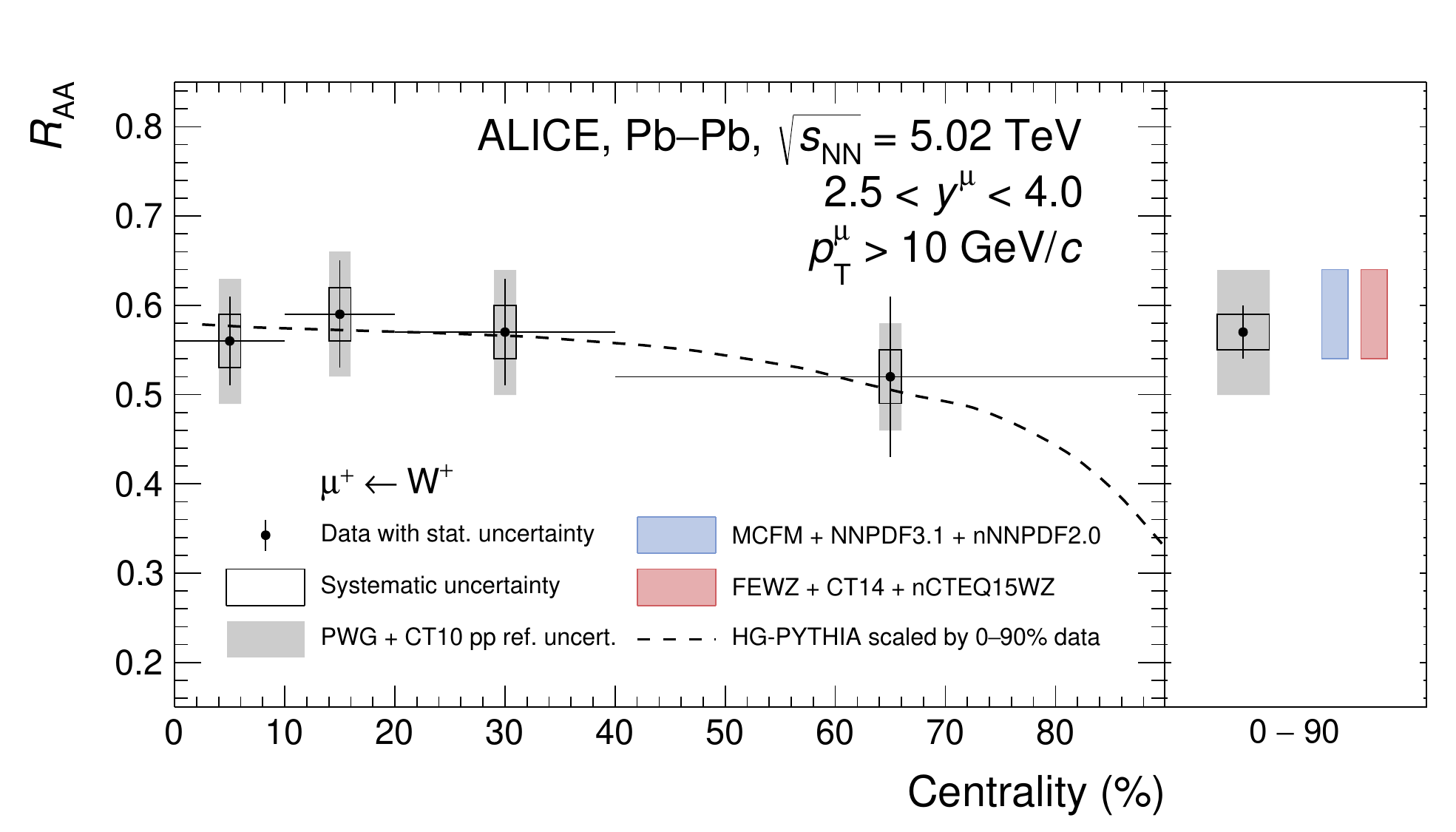}
    \end{center}
    \caption{Nuclear modification factor of muons from \Wminus (top) and \Wplus (bottom) decays in \PbPb collisions at \fivenn, for muons with $\pt^\mu > 10$ \GeVc and $2.5 < y^\mu_{\rm cms} < 4.0$, in different centrality intervals (left panels) and for the 0--90\% range (right panels). The centrality-dependent distributions are compared with the dashed curve, corresponding to the HG-PYTHIA~\cite{hg-pythia} model scaled with the measured $R_{\rm AA}$ in the 0--90\% centrality interval. The centrality-averaged measurement is compared with CT14+nCTEQ15WZ~\cite{ct14,ncteq15wz} and NNPDF3.1+nNNPDF2.0~\cite{nnpdf,nnnpdf} calculations. The horizontal bars indicate the width of the centrality bins, the vertical bars and boxes correspond the statistical and systematic uncertainties on the \PbPb measurement, respectively. The grey boxes indicate the uncertainty on the pp reference cross section.}
    \label{fig:PbPb_Raa}
\end{figure}

The measured \RPbPb is shown in Fig.~\ref{fig:PbPb_Raa} as a function of centrality and for the 0--90\% centrality class. The centrality-dependent measurement is compared with HG-PYTHIA~\cite{hg-pythia} calculations of the \RPbPb of hard scatterings scaled with the measured value in 0--90\% centrality. The centrality-averaged \RPbPb is compared with pQCD calculations, using the CT14~\cite{ct14} PDFs for the proton and the nCTEQ15WZ~\cite{ncteq15wz} PDF set, or the NNPDF3.1+nNNPDF2.0 combination~\cite{nnpdf,nnnpdf} for the Pb nucleus. The calculations within the nCTEQ and NNPDF frameworks are only shown for the centrality-averaged value as they have no centrality dependence. Both models provide a good description of the measurement within uncertainties. It should be noted that this agreement is realised while the measurement and models use different PDF sets for the pp reference, and different codes for the pQCD calculations (POWHEG~\cite{powheg} for the experimental results, MCFM~\cite{mcfm} and FEWZ~\cite{fewz} for the theoretical ones).

\section{Summary}
\label{sec:summary}

The measurements of the \W-boson production cross section, lepton charge asymmetry, nuclear modification factor, and yield normalised to the number of nucleon--nucleon collisions in \pPb collisions at \eightnn were reported, constituting the first results on the production of \W bosons at large rapidity at this energy, extending the measurement in \pPb collisions at \fivenn with a significant improvement of the precision. They were performed for muons with $\pt^\mu >$ 10 \GeVc and in the rapidity intervals $-4.46 < y^\mu_{\rm cms} < -2.96$ and $2.03 < y^\mu_{\rm cms} < 3.53$, where the negative rapidity interval indicates the Pb-going side, and the positive one the p-going side. The results were compared with pQCD calculations, using the CT14 PDF set~\cite{ct14}, and the EPPS16~\cite{epps16}, nNNPDF2.0~\cite{nnnpdf} and nCTEQ15~\cite{ncteq15, ncteq15wz} nPDF models. Some tensions are observed in the ability of the models to reproduce the data, notably in the rapidity dependence of the observables. Significant deviations from the free-nucleon PDF predictions, up to 3.5$\sigma$, are found at forward rapidity, corresponding to the shadowing region of the nuclear modifications at low Bjorken-$x$. The measurements in \pPb collisions reported here can therefore provide significant constraints to the nPDF models and help reducing their uncertainties. They complement the measurements of the \Z-boson production performed at large rapidities by the ALICE Collaboration~\cite{Alice:ZpPb8tevPbPb5tev}, where the statistical precision was too limited to draw any conclusion on the nuclear modifications. The comparison with the CMS measurements at midrapidity illustrates the complementarity of the LHC experiments in providing such results. The binary scaling of hard processes is observed, as the production cross sections in different centrality classes normalised to the average number of binary nucleon--nucleon collisions were found to be constant within uncertainties.

Similar measurements performed in \PbPb collisions at \fivenn were also presented, for muons from \W-boson decays at large rapidity ($2.5 < y^\mu_{\rm cms} < 4.0$) with $\pt^\mu >$ 10 \GeVc, and for various centrality classes. The normalised yield as a function of centrality follows the binary scaling expected for a hard process in the absence of significant centrality dependence of the shadowing. Comparisons with pQCD calculations show the ability of the EPPS16, nCTEQ15WZ and nNNPDF2.0 nPDFs to reproduce the production cross section, lepton charge asymmetry and nuclear modification factor. The evaluation of the \avTaa-normalised yield with the nuclear-suppressed inelastic nucleon--nucleon cross section $\sigma^{\rm inel}_{\rm NN}$ obtained from Ref.~\cite{epps-new}, which was found to improve the agreement between the ATLAS data and the EPPS16 model, yields a tension between the ALICE measurements and EPPS16 calculations for peripheral events which is not seen with the standard value of $\sigma^{\rm inel}_{\rm NN}$. The \avTaa-scaled yield and the nuclear modification factor are found to be in good agreement with HG-PYTHIA~\cite{hg-pythia} calculations of the \RPbPb of hard scatterings scaled with the value measured in 0--90\% centrality, but the statistical limitation of the measurement does not allow to conclude on the decrease in peripheral events expected from this model. The measured \avTaa-scaled yields are described by pQCD calculations with the EPPS16 nPDFs. These measurements support the conclusion derived from the measurement of the \Z-boson production in \PbPb collisions at \fivenn~\cite{Alice:ZpPb8tevPbPb5tev}, showing a suppression of the production of electroweak bosons due to the nuclear modifications of the PDF and the resulting deviations from calculations based on free-nucleon PDFs. Being the first measurement of the \W production in \PbPb collisions at large rapidity, this study provides important insights for further investigation of the centrality dependence of the nPDFs.


\newenvironment{acknowledgement}{\relax}{\relax}
\begin{acknowledgement}
\section*{Acknowledgements}
The authors would like to thank Hannu Paukkunen, Rabah Abdul Khalek and Aleksander Kusina for providing the various model calculations.

The ALICE Collaboration would like to thank all its engineers and technicians for their invaluable contributions to the construction of the experiment and the CERN accelerator teams for the outstanding performance of the LHC complex.
The ALICE Collaboration gratefully acknowledges the resources and support provided by all Grid centres and the Worldwide LHC Computing Grid (WLCG) collaboration.
The ALICE Collaboration acknowledges the following funding agencies for their support in building and running the ALICE detector:
A. I. Alikhanyan National Science Laboratory (Yerevan Physics Institute) Foundation (ANSL), State Committee of Science and World Federation of Scientists (WFS), Armenia;
Austrian Academy of Sciences, Austrian Science Fund (FWF): [M 2467-N36] and Nationalstiftung f\"{u}r Forschung, Technologie und Entwicklung, Austria;
Ministry of Communications and High Technologies, National Nuclear Research Center, Azerbaijan;
Conselho Nacional de Desenvolvimento Cient\'{\i}fico e Tecnol\'{o}gico (CNPq), Financiadora de Estudos e Projetos (Finep), Funda\c{c}\~{a}o de Amparo \`{a} Pesquisa do Estado de S\~{a}o Paulo (FAPESP) and Universidade Federal do Rio Grande do Sul (UFRGS), Brazil;
Bulgarian Ministry of Education and Science, within the National Roadmap for Research Infrastructures 2020¿2027 (object CERN), Bulgaria;
Ministry of Education of China (MOEC) , Ministry of Science \& Technology of China (MSTC) and National Natural Science Foundation of China (NSFC), China;
Ministry of Science and Education and Croatian Science Foundation, Croatia;
Centro de Aplicaciones Tecnol\'{o}gicas y Desarrollo Nuclear (CEADEN), Cubaenerg\'{\i}a, Cuba;
Ministry of Education, Youth and Sports of the Czech Republic, Czech Republic;
The Danish Council for Independent Research | Natural Sciences, the VILLUM FONDEN and Danish National Research Foundation (DNRF), Denmark;
Helsinki Institute of Physics (HIP), Finland;
Commissariat \`{a} l'Energie Atomique (CEA) and Institut National de Physique Nucl\'{e}aire et de Physique des Particules (IN2P3) and Centre National de la Recherche Scientifique (CNRS), France;
Bundesministerium f\"{u}r Bildung und Forschung (BMBF) and GSI Helmholtzzentrum f\"{u}r Schwerionenforschung GmbH, Germany;
General Secretariat for Research and Technology, Ministry of Education, Research and Religions, Greece;
National Research, Development and Innovation Office, Hungary;
Department of Atomic Energy Government of India (DAE), Department of Science and Technology, Government of India (DST), University Grants Commission, Government of India (UGC) and Council of Scientific and Industrial Research (CSIR), India;
National Research and Innovation Agency - BRIN, Indonesia;
Istituto Nazionale di Fisica Nucleare (INFN), Italy;
Japanese Ministry of Education, Culture, Sports, Science and Technology (MEXT) and Japan Society for the Promotion of Science (JSPS) KAKENHI, Japan;
Consejo Nacional de Ciencia (CONACYT) y Tecnolog\'{i}a, through Fondo de Cooperaci\'{o}n Internacional en Ciencia y Tecnolog\'{i}a (FONCICYT) and Direcci\'{o}n General de Asuntos del Personal Academico (DGAPA), Mexico;
Nederlandse Organisatie voor Wetenschappelijk Onderzoek (NWO), Netherlands;
The Research Council of Norway, Norway;
Commission on Science and Technology for Sustainable Development in the South (COMSATS), Pakistan;
Pontificia Universidad Cat\'{o}lica del Per\'{u}, Peru;
Ministry of Education and Science, National Science Centre and WUT ID-UB, Poland;
Korea Institute of Science and Technology Information and National Research Foundation of Korea (NRF), Republic of Korea;
Ministry of Education and Scientific Research, Institute of Atomic Physics, Ministry of Research and Innovation and Institute of Atomic Physics and University Politehnica of Bucharest, Romania;
Ministry of Education, Science, Research and Sport of the Slovak Republic, Slovakia;
National Research Foundation of South Africa, South Africa;
Swedish Research Council (VR) and Knut \& Alice Wallenberg Foundation (KAW), Sweden;
European Organization for Nuclear Research, Switzerland;
Suranaree University of Technology (SUT), National Science and Technology Development Agency (NSTDA), Thailand Science Research and Innovation (TSRI) and National Science, Research and Innovation Fund (NSRF), Thailand;
Turkish Energy, Nuclear and Mineral Research Agency (TENMAK), Turkey;
National Academy of  Sciences of Ukraine, Ukraine;
Science and Technology Facilities Council (STFC), United Kingdom;
National Science Foundation of the United States of America (NSF) and United States Department of Energy, Office of Nuclear Physics (DOE NP), United States of America.
In addition, individual groups or members have received support from:
Marie Sk\l{}odowska Curie, Strong 2020 - Horizon 2020, European Research Council (grant nos. 824093, 896850, 950692), European Union;
Academy of Finland (Center of Excellence in Quark Matter) (grant nos. 346327, 346328), Finland;
Programa de Apoyos para la Superaci\'{o}n del Personal Acad\'{e}mico, UNAM, Mexico.
\end{acknowledgement}

\bibliographystyle{utphys}   
\bibliography{bibliography}

\newpage
\appendix

%
%

\section{The ALICE Collaboration}
\label{app:collab}
\begin{flushleft} 
\small

S.~Acharya\,\orcidlink{0000-0002-9213-5329}\,$^{\rm 124,131}$, 
D.~Adamov\'{a}\,\orcidlink{0000-0002-0504-7428}\,$^{\rm 86}$, 
A.~Adler$^{\rm 69}$, 
G.~Aglieri Rinella\,\orcidlink{0000-0002-9611-3696}\,$^{\rm 32}$, 
M.~Agnello\,\orcidlink{0000-0002-0760-5075}\,$^{\rm 29}$, 
N.~Agrawal\,\orcidlink{0000-0003-0348-9836}\,$^{\rm 50}$, 
Z.~Ahammed\,\orcidlink{0000-0001-5241-7412}\,$^{\rm 131}$, 
S.~Ahmad\,\orcidlink{0000-0003-0497-5705}\,$^{\rm 15}$, 
S.U.~Ahn\,\orcidlink{0000-0001-8847-489X}\,$^{\rm 70}$, 
I.~Ahuja\,\orcidlink{0000-0002-4417-1392}\,$^{\rm 37}$
% ...

\section*{Affiliation Notes}

$^{\rm I}$ Deceased\\
$^{\rm II}$ Also at: Max-Planck-Institut f\"{u}r Physik, Munich, Germany\\
$^{\rm III}$ Also at: Italian National Agency for New Technologies, Energy and Sustainable Economic Development (ENEA), Bologna, Italy\\
$^{\rm IV}$ Also at: Dipartimento DET del Politecnico di Torino, Turin, Italy\\
$^{\rm V}$ Also at: Department of Applied Physics, Aligarh Muslim University, Aligarh, India\\
$^{\rm VI}$ Also at: Institute of Theoretical Physics, University of Wroclaw, Poland\\
$^{\rm VII}$ Also at: An institution covered by a cooperation agreement with CERN\\

\section*{Collaboration Institutes}

$^{1}$ A.I. Alikhanyan National Science Laboratory (Yerevan Physics Institute) Foundation, Yerevan, Armenia\\
$^{2}$ AGH University of Science and Technology, Cracow, Poland\\
$^{3}$ Bogolyubov Institute for Theoretical Physics, National Academy of Sciences of Ukraine, Kiev, Ukraine\\
$^{4}$ Bose Institute, Department of Physics  and Centre for Astroparticle Physics and Space Science (CAPSS), Kolkata, India\\
$^{5}$ California Polytechnic State University, San Luis Obispo, California, United States\\
$^{6}$ Central China Normal University, Wuhan, China\\
$^{7}$ Centro de Aplicaciones Tecnol\'{o}gicas y Desarrollo Nuclear (CEADEN), Havana, Cuba\\
$^{8}$ Centro de Investigaci\'{o}n y de Estudios Avanzados (CINVESTAV), Mexico City and M\'{e}rida, Mexico\\
$^{9}$ Chicago State University, Chicago, Illinois, United States\\
$^{10}$ China Institute of Atomic Energy, Beijing, China\\
$^{11}$ Chungbuk National University, Cheongju, Republic of Korea\\
$^{12}$ Comenius University Bratislava, Faculty of Mathematics, Physics and Informatics, Bratislava, Slovak Republic\\
$^{13}$ COMSATS University Islamabad, Islamabad, Pakistan\\
$^{14}$ Creighton University, Omaha, Nebraska, United States\\
$^{15}$ Department of Physics, Aligarh Muslim University, Aligarh, India\\
$^{16}$ Department of Physics, Pusan National University, Pusan, Republic of Korea\\
$^{17}$ Department of Physics, Sejong University, Seoul, Republic of Korea\\
$^{18}$ Department of Physics, University of California, Berkeley, California, United States\\
$^{19}$ Department of Physics, University of Oslo, Oslo, Norway\\
$^{20}$ Department of Physics and Technology, University of Bergen, Bergen, Norway\\
$^{21}$ Dipartimento di Fisica, Universit\`{a} di Pavia, Pavia, Italy\\
$^{22}$ Dipartimento di Fisica dell'Universit\`{a} and Sezione INFN, Cagliari, Italy\\
$^{23}$ Dipartimento di Fisica dell'Universit\`{a} and Sezione INFN, Trieste, Italy\\
$^{24}$ Dipartimento di Fisica dell'Universit\`{a} and Sezione INFN, Turin, Italy\\
$^{25}$ Dipartimento di Fisica e Astronomia dell'Universit\`{a} and Sezione INFN, Bologna, Italy\\
$^{26}$ Dipartimento di Fisica e Astronomia dell'Universit\`{a} and Sezione INFN, Catania, Italy\\
$^{27}$ Dipartimento di Fisica e Astronomia dell'Universit\`{a} and Sezione INFN, Padova, Italy\\
$^{28}$ Dipartimento di Fisica `E.R.~Caianiello' dell'Universit\`{a} and Gruppo Collegato INFN, Salerno, Italy\\
$^{29}$ Dipartimento DISAT del Politecnico and Sezione INFN, Turin, Italy\\
$^{30}$ Dipartimento di Scienze MIFT, Universit\`{a} di Messina, Messina, Italy\\
$^{31}$ Dipartimento Interateneo di Fisica `M.~Merlin' and Sezione INFN, Bari, Italy\\
$^{32}$ European Organization for Nuclear Research (CERN), Geneva, Switzerland\\
$^{33}$ Faculty of Electrical Engineering, Mechanical Engineering and Naval Architecture, University of Split, Split, Croatia\\
$^{34}$ Faculty of Engineering and Science, Western Norway University of Applied Sciences, Bergen, Norway\\
$^{35}$ Faculty of Nuclear Sciences and Physical Engineering, Czech Technical University in Prague, Prague, Czech Republic\\
$^{36}$ Faculty of Physics, Sofia University, Sofia, Bulgaria\\
$^{37}$ Faculty of Science, P.J.~\v{S}af\'{a}rik University, Ko\v{s}ice, Slovak Republic\\
$^{38}$ Frankfurt Institute for Advanced Studies, Johann Wolfgang Goethe-Universit\"{a}t Frankfurt, Frankfurt, Germany\\
$^{39}$ Fudan University, Shanghai, China\\
$^{40}$ Gangneung-Wonju National University, Gangneung, Republic of Korea\\
$^{41}$ Gauhati University, Department of Physics, Guwahati, India\\
$^{42}$ Helmholtz-Institut f\"{u}r Strahlen- und Kernphysik, Rheinische Friedrich-Wilhelms-Universit\"{a}t Bonn, Bonn, Germany\\
$^{43}$ Helsinki Institute of Physics (HIP), Helsinki, Finland\\
$^{44}$ High Energy Physics Group,  Universidad Aut\'{o}noma de Puebla, Puebla, Mexico\\
$^{45}$ Horia Hulubei National Institute of Physics and Nuclear Engineering, Bucharest, Romania\\
$^{46}$ Indian Institute of Technology Bombay (IIT), Mumbai, India\\
$^{47}$ Indian Institute of Technology Indore, Indore, India\\
$^{48}$ INFN, Laboratori Nazionali di Frascati, Frascati, Italy\\
$^{49}$ INFN, Sezione di Bari, Bari, Italy\\
$^{50}$ INFN, Sezione di Bologna, Bologna, Italy\\
$^{51}$ INFN, Sezione di Cagliari, Cagliari, Italy\\
$^{52}$ INFN, Sezione di Catania, Catania, Italy\\
$^{53}$ INFN, Sezione di Padova, Padova, Italy\\
$^{54}$ INFN, Sezione di Pavia, Pavia, Italy\\
$^{55}$ INFN, Sezione di Torino, Turin, Italy\\
$^{56}$ INFN, Sezione di Trieste, Trieste, Italy\\
$^{57}$ Inha University, Incheon, Republic of Korea\\
$^{58}$ Institute for Gravitational and Subatomic Physics (GRASP), Utrecht University/Nikhef, Utrecht, Netherlands\\
$^{59}$ Institute of Experimental Physics, Slovak Academy of Sciences, Ko\v{s}ice, Slovak Republic\\
$^{60}$ Institute of Physics, Homi Bhabha National Institute, Bhubaneswar, India\\
$^{61}$ Institute of Physics of the Czech Academy of Sciences, Prague, Czech Republic\\
$^{62}$ Institute of Space Science (ISS), Bucharest, Romania\\
$^{63}$ Institut f\"{u}r Kernphysik, Johann Wolfgang Goethe-Universit\"{a}t Frankfurt, Frankfurt, Germany\\
$^{64}$ Instituto de Ciencias Nucleares, Universidad Nacional Aut\'{o}noma de M\'{e}xico, Mexico City, Mexico\\
$^{65}$ Instituto de F\'{i}sica, Universidade Federal do Rio Grande do Sul (UFRGS), Porto Alegre, Brazil\\
$^{66}$ Instituto de F\'{\i}sica, Universidad Nacional Aut\'{o}noma de M\'{e}xico, Mexico City, Mexico\\
$^{67}$ iThemba LABS, National Research Foundation, Somerset West, South Africa\\
$^{68}$ Jeonbuk National University, Jeonju, Republic of Korea\\
$^{69}$ Johann-Wolfgang-Goethe Universit\"{a}t Frankfurt Institut f\"{u}r Informatik, Fachbereich Informatik und Mathematik, Frankfurt, Germany\\
$^{70}$ Korea Institute of Science and Technology Information, Daejeon, Republic of Korea\\
$^{71}$ KTO Karatay University, Konya, Turkey\\
$^{72}$ Laboratoire de Physique des 2 Infinis, Ir\`{e}ne Joliot-Curie, Orsay, France\\
$^{73}$ Laboratoire de Physique Subatomique et de Cosmologie, Universit\'{e} Grenoble-Alpes, CNRS-IN2P3, Grenoble, France\\
$^{74}$ Lawrence Berkeley National Laboratory, Berkeley, California, United States\\
$^{75}$ Lund University Department of Physics, Division of Particle Physics, Lund, Sweden\\
$^{76}$ Nagasaki Institute of Applied Science, Nagasaki, Japan\\
$^{77}$ Nara Women{'}s University (NWU), Nara, Japan\\
$^{78}$ National and Kapodistrian University of Athens, School of Science, Department of Physics , Athens, Greece\\
$^{79}$ National Centre for Nuclear Research, Warsaw, Poland\\
$^{80}$ National Institute of Science Education and Research, Homi Bhabha National Institute, Jatni, India\\
$^{81}$ National Nuclear Research Center, Baku, Azerbaijan\\
$^{82}$ National Research and Innovation Agency - BRIN, Jakarta, Indonesia\\
$^{83}$ Niels Bohr Institute, University of Copenhagen, Copenhagen, Denmark\\
$^{84}$ Nikhef, National institute for subatomic physics, Amsterdam, Netherlands\\
$^{85}$ Nuclear Physics Group, STFC Daresbury Laboratory, Daresbury, United Kingdom\\
$^{86}$ Nuclear Physics Institute of the Czech Academy of Sciences, Husinec-\v{R}e\v{z}, Czech Republic\\
$^{87}$ Oak Ridge National Laboratory, Oak Ridge, Tennessee, United States\\
$^{88}$ Ohio State University, Columbus, Ohio, United States\\
$^{89}$ Physics department, Faculty of science, University of Zagreb, Zagreb, Croatia\\
$^{90}$ Physics Department, Panjab University, Chandigarh, India\\
$^{91}$ Physics Department, University of Jammu, Jammu, India\\
$^{92}$ Physics Department, University of Rajasthan, Jaipur, India\\
$^{93}$ Physics Program and International Institute for Sustainability with Knotted Chiral Meta Matter (SKCM2), Hiroshima University, Hiroshima, Japan\\
$^{94}$ Physikalisches Institut, Eberhard-Karls-Universit\"{a}t T\"{u}bingen, T\"{u}bingen, Germany\\
$^{95}$ Physikalisches Institut, Ruprecht-Karls-Universit\"{a}t Heidelberg, Heidelberg, Germany\\
$^{96}$ Physik Department, Technische Universit\"{a}t M\"{u}nchen, Munich, Germany\\
$^{97}$ Politecnico di Bari and Sezione INFN, Bari, Italy\\
$^{98}$ Research Division and ExtreMe Matter Institute EMMI, GSI Helmholtzzentrum f\"ur Schwerionenforschung GmbH, Darmstadt, Germany\\
$^{99}$ Saha Institute of Nuclear Physics, Homi Bhabha National Institute, Kolkata, India\\
$^{100}$ School of Physics and Astronomy, University of Birmingham, Birmingham, United Kingdom\\
$^{101}$ Secci\'{o}n F\'{\i}sica, Departamento de Ciencias, Pontificia Universidad Cat\'{o}lica del Per\'{u}, Lima, Peru\\
$^{102}$ Stefan Meyer Institut f\"{u}r Subatomare Physik (SMI), Vienna, Austria\\
$^{103}$ SUBATECH, IMT Atlantique, Nantes Universit\'{e}, CNRS-IN2P3, Nantes, France\\
$^{104}$ Suranaree University of Technology, Nakhon Ratchasima, Thailand\\
$^{105}$ Technical University of Ko\v{s}ice, Ko\v{s}ice, Slovak Republic\\
$^{106}$ The Henryk Niewodniczanski Institute of Nuclear Physics, Polish Academy of Sciences, Cracow, Poland\\
$^{107}$ The University of Texas at Austin, Austin, Texas, United States\\
$^{108}$ Universidad Aut\'{o}noma de Sinaloa, Culiac\'{a}n, Mexico\\
$^{109}$ Universidade de S\~{a}o Paulo (USP), S\~{a}o Paulo, Brazil\\
$^{110}$ Universidade Estadual de Campinas (UNICAMP), Campinas, Brazil\\
$^{111}$ Universidade Federal do ABC, Santo Andre, Brazil\\
$^{112}$ University of Cape Town, Cape Town, South Africa\\
$^{113}$ University of Houston, Houston, Texas, United States\\
$^{114}$ University of Jyv\"{a}skyl\"{a}, Jyv\"{a}skyl\"{a}, Finland\\
$^{115}$ University of Kansas, Lawrence, Kansas, United States\\
$^{116}$ University of Liverpool, Liverpool, United Kingdom\\
$^{117}$ University of Science and Technology of China, Hefei, China\\
$^{118}$ University of South-Eastern Norway, Kongsberg, Norway\\
$^{119}$ University of Tennessee, Knoxville, Tennessee, United States\\
$^{120}$ University of the Witwatersrand, Johannesburg, South Africa\\
$^{121}$ University of Tokyo, Tokyo, Japan\\
$^{122}$ University of Tsukuba, Tsukuba, Japan\\
$^{123}$ University Politehnica of Bucharest, Bucharest, Romania\\
$^{124}$ Universit\'{e} Clermont Auvergne, CNRS/IN2P3, LPC, Clermont-Ferrand, France\\
$^{125}$ Universit\'{e} de Lyon, CNRS/IN2P3, Institut de Physique des 2 Infinis de Lyon, Lyon, France\\
$^{126}$ Universit\'{e} de Strasbourg, CNRS, IPHC UMR 7178, F-67000 Strasbourg, France, Strasbourg, France\\
$^{127}$ Universit\'{e} Paris-Saclay Centre d'Etudes de Saclay (CEA), IRFU, D\'{e}partment de Physique Nucl\'{e}aire (DPhN), Saclay, France\\
$^{128}$ Universit\`{a} degli Studi di Foggia, Foggia, Italy\\
$^{129}$ Universit\`{a} del Piemonte Orientale, Vercelli, Italy\\
$^{130}$ Universit\`{a} di Brescia, Brescia, Italy\\
$^{131}$ Variable Energy Cyclotron Centre, Homi Bhabha National Institute, Kolkata, India\\
$^{132}$ Warsaw University of Technology, Warsaw, Poland\\
$^{133}$ Wayne State University, Detroit, Michigan, United States\\
$^{134}$ Westf\"{a}lische Wilhelms-Universit\"{a}t M\"{u}nster, Institut f\"{u}r Kernphysik, M\"{u}nster, Germany\\
$^{135}$ Wigner Research Centre for Physics, Budapest, Hungary\\
$^{136}$ Yale University, New Haven, Connecticut, United States\\
$^{137}$ Yonsei University, Seoul, Republic of Korea\\
$^{138}$  Zentrum  f\"{u}r Technologie und Transfer (ZTT), Worms, Germany\\
$^{139}$ Affiliated with an institute covered by a cooperation agreement with CERN\\
$^{140}$ Affiliated with an international laboratory covered by a cooperation agreement with CERN.\\

\end{flushleft} 

\end{document}